\pdfoutput=1
\RequirePackage{ifpdf}
\ifpdf 
\documentclass[pdftex]{sigma}
\else
\documentclass{sigma}
\fi

\graphicspath{{figures/}}

\numberwithin{equation}{section}

\newtheorem*{Theorem*}{Theorem}

 { \theoremstyle{definition}

 }

\newcommand{\CC}{{\cal C}}

\newcommand{\CF}{{\cal F}}
\newcommand{\CG}{{\cal G}}

\newcommand{\CI}{{\cal I}}

\newcommand{\CL}{{\cal L}}

\newcommand{\CO}{{\cal O}}

\def\IZ{{\mathbb Z}}
\def\IR{{\mathbb R}}
\def\IC{{\mathbb C}}
\def\IN{{\mathbb N}}

\newcommand{\mQ}{\mathsf{Q}}

\DeclareMathOperator{\Res}{Res}

\DeclareMathOperator{\Tr}{Tr}
\DeclareMathOperator{\coeff}{coeff}

\newcommand{\re}{{\rm e}}
\newcommand{\ri}{{\rm i}}
\newcommand{\rd}{{\rm d}}

\newcommand{\be}{\begin{equation}}
\newcommand{\ee}{\end{equation}}
\newcommand{\ba}{\begin{aligned}}
\newcommand{\ea}{\end{aligned}}

\newdimen\tableauside\tableauside=1.0ex
\newdimen\tableaurule\tableaurule=0.4pt
\newdimen\tableaustep
\def\phantomhrule#1{\hbox{\vbox to0pt{\hrule height\tableaurule width#1\vss}}}
\def\phantomvrule#1{\vbox{\hbox to0pt{\vrule width\tableaurule height#1\hss}}}
\def\sqr{\vbox{%
 \phantomhrule\tableaustep
 \hbox{\phantomvrule\tableaustep\kern\tableaustep\phantomvrule\tableaustep}%
 \hbox{\vbox{\phantomhrule\tableauside}\kern-\tableaurule}}}
\def\squares#1{\hbox{\count0=#1\noindent\loop\sqr
 \advance\count0 by-1 \ifnum\count0>0\repeat}}
\def\tableau#1{\vcenter{\offinterlineskip
 \tableaustep=\tableauside\advance\tableaustep by-\tableaurule
 \kern\normallineskip\hbox
 {\kern\normallineskip\vbox
 {\gettableau#1 0 }%
 \kern\normallineskip\kern\tableaurule}%
 \kern\normallineskip\kern\tableaurule}}
\def\gettableau#1{\ifnum#1=0\let\next=\null\else
\squares{#1}\let\next=\gettableau\fi\next}

\usepackage{mathtools}
\usepackage{slashed}
\usepackage{booktabs}

\begin{document}

\newcommand{\arXivNumber}{2302.08363}

\renewcommand{\PaperNumber}{065}

\FirstPageHeading

\ShortArticleName{On the Structure of Trans-Series in Quantum Field Theory}

\ArticleName{On the Structure of Trans-Series\\ in Quantum Field Theory}

\Author{Marcos MARI\~NO~$^{\rm a}$, Ramon MIRAVITLLAS~$^{\rm ab}$ and Tom\'as REIS~$^{\rm cd}$}

\AuthorNameForHeading{M.~Mari\~no, R.~Miravitllas and T.~Reis}

\Address{$^{\rm a)}$~D\'epartement de Physique Th\'eorique et Section de Math\'ematiques,\\
\hphantom{$^{\rm a)}$}~Universit\'e de Gen\`eve, Gen\`eve, CH-1211, Switzerland}
\EmailD{\href{mailto:marcos.marino@unige.ch}{marcos.marino@unige.ch}}

\Address{$^{\rm b)}$~HUN-REN Wigner Research Centre for Physics, \\
\hphantom{$^{\rm b)}$}~Konkoly-Thege Mikl\'os u. 29-33, 1121 Budapest, Hungary}
\EmailD{\href{mailto:ramon.miravitllas.mas@wigner.hu}{ramon.miravitllas.mas@wigner.hu}}

\Address{$^{\rm c)}$~SISSA, 34136 Trieste, Italy}
\EmailD{\href{mailto:treis@sissa.it}{treis@sissa.it}}

\Address{$^{\rm d)}$~INFN, Sezione di Trieste, 34127 Trieste, Italy}

\ArticleDates{Received March 06, 2025, in final form July 22, 2025; Published online August 01, 2025}

\Abstract{Many observables in quantum field theory can be expressed in~terms of trans-series, in which one adds to the perturbative series a typically infinite sum of exponentially small corrections, due to instantons or to renormalons. Even after Borel resummation of the series in~the coupling constant, one has to sum this infinite series of small exponential corrections. It has been argued that this leads to a new divergence, which is sometimes called the divergence of the OPE. We show that, in some interesting examples in quantum field theory, the series of small exponential corrections is convergent, order by order in~the coupling constant. In particular, we give numerical evidence for this convergence property in~the case of the free energy of integrable asymptotically free theories, which has been intensively studied recently in~the framework of resurgence. Our results indicate that, in~these examples, the Borel resummed trans-series leads to a well defined function, and there are no further divergences.}


\Classification{81T40; 40G10; 30E15}

\section{Introduction}

 It is widely believed that renormalized perturbation theory gives the asymptotic expansions of observables in quantum field
 theory (QFT), when the coupling is small. The perturbative series obtained in~this way are typically factorially divergent,
 therefore they can only give approximate results for the observables. However, there are many situations
 in mathematics and physics where perturbative series can be upgraded to {\it trans-series}, which provide exponentially small
 corrections to the perturbative series. These
 trans-series then lead to exact results by using
 Borel resummation. An example where this happens is the theory of non-linear ODEs developed by \'Ecalle,
 Costin and others (see~\cite{bssv, costin, costin2, costin-costin, ecalle}, and~\cite{ss} for a very readable introduction).
 Another example is the exact WKB method in quantum mechanics, where an exact version of the Bohr--Sommerfeld
 quantization condition is obtained by resumming a trans-series~\mbox{\cite{ddpham,power, voros}}.\looseness=1

 It has been conjectured in various forms and
occasions that perturbative series in QFT can be also upgraded to
trans-series, in such a way that exact results can be obtained by their resummation, in some domain
of convergence of the coupling constant. Proving such a~statement in a non-trivial QFT is rather difficult, and even
checking it (say, numerically) is not easy, since there are very
few examples in QFT where one can compute the actual trans-series reliably.\looseness=1

 The resummation of a trans-series has, roughly speaking, two steps. First, one resums each
 factorially divergent power series in~the coupling constant, associated to a given exponential correction. Then, one has to sum
 over all the exponentially small corrections. This last sum is typically over an infinite number of terms,
 and its convergence (or lack thereof) poses an~additional problem. In the case of non-linear ODEs, both steps can be performed, and the sum over
 exponentially small corrections converges to the actual solution of the ODE, provided the expansion parameter is not too large~\cite{costin,costin2, ecalle}. In the case of generic observables in QFT, it has been argued
~\cite{shifman,shifman-hadrons, shifman-renormalons} that the sum of exponentially
 small corrections is generically factorially divergent. In particular, it does not hold in~the case of two-point functions in asymptotically free
theories, like the Adler function in QCD. The reason is that the singularity structure of general
correlation functions cannot be reproduced with a convergent trans-series. This phenomenon is sometimes called, in~the context of
renormalon physics, the divergence of the OPE, and it is at the basis of the so-called violation
of quark-hadron duality (see~\cite{peris} for a recent review). In~these cases,
the reconstruction of the exact answer by Borel resummation of the trans-series is more involved,
since one has to do a further resummation of the factorially divergent series of exponentially small corrections.

In this paper, we want to explore a property of trans-series, inspired by the theory of non-linear ODEs,
which provides indirect evidence for the convergence (or not) of the trans-series. In a trans-series expansion we have typically
two small expansion parameters: the coupling constant $g$, and an appropriate exponential thereof
\smash{$\re^{-A/g}$}. At a given order in~the exponential, the resulting series in $g$ is typically
factorially divergent. However, one can
consider the opposite regime: at each order in $g$, one has a series in
\smash{$\re^{-A/g}$}, which we will simply call partial series. In~the theory of non-linear ODEs, it can be proved
\cite{costin-costin} that all the partial series are {\it convergent}, and with the {\it same} radius of convergence. We will
call this property the convergence property of partial series. This property also holds in~the trans-series appearing naturally
in~the exact WKB method in quantum mechanics, like, e.g., in~the
Delabaere--Pham formula~\cite{dpham, voros-quartic} and in~the exact quantization conditions of~\cite{ddpham, voros}.
In this paper we show that the convergence of the partial series also
holds in some interesting examples in QFT. In particular, we give evidence that it holds for
an observable much studied recently from the point of view of resurgence, namely the free
energy of integrable, asymptotically free 2d QFTs~\cite{abbh1,
abbh2, bbhv, bbhv2,bbv,bbv2,dmss,fkw-letter,fkw,mmr, mmr-an, mmr-theta, mr-ren,s-thompson, volin}. It is expected that the convergence property of the partial series
leads to the convergence of the full resummed trans-series, as it happens in~the case of ODEs~\cite{costin-costin}.
The advantage of the convergence
property is that it can be addressed directly in~the formal trans-series, before
Borel resummation. The evidence we give for the convergence property in~the case of the
free energy shows that this observable behaves very differently from correlation functions. In particular,
its singularity structure must be qualitatively different from what is found in, e.g., typical two-point functions.\looseness=1

This paper is organized as follows. In Section~\ref{tss-sec}, we present the basic
structure of trans-series in quantum field theory, and we
define the partial series collecting the different series of exponentially small corrections. In Section~\ref{sec-sigma}, we
look at a simple example where the convergence of partial series can be tested in detail, namely the two-point function
of the large $N$ sigma model. In Section~\ref{sec-cs}, we consider a similar simple example in complex Chern--Simons theory.
The core of the paper is contained in Section~\ref{sec-integrable}, where we test the convergence of partial series in
the case of the free energy of integrable models. Finally, we conclude with a summary of the results and prospects for
future developments.

\section{Trans-series and their structure} \label{tss-sec}
In this paper, we will consider trans-series involving two small parameters: a coupling
constant~$g$ and an exponentially small term of the form
\be
y = g^{-b} \re^{-A/g}.
\ee
These trans-series are of the form
\be
\label{phits}
\Phi= \sum_{\ell \ge 0} y^{\ell} \varphi_\ell (g),
\ee
where
\be
 \label{trans-not}
 \varphi_\ell (g)= \sum_{n \ge 0} a_{n, \ell} g^n
\ee
 are formal power series in $g$. Some of the examples we will consider
 will be however slightly more general, including for example logarithmic terms, or involving
 finite sums of expressions like~(\ref{phits}). The series $\varphi_0(g)$ is the perturbative series,
 and the higher order series $\varphi_\ell (g)$, with~${\ell\ge 1}$,
 are non-perturbative corrections. These could be of the instanton or the renormalon type,
and in~this paper we will consider both of them. In general the coefficients $a_{n, \ell}$ for $\ell\ge 1$ are complex and they
depend on a choice of lateral resummation (in~the case of ODEs, there is a~dependence on a trans-series parameter $C$,
and different choices of resummation lead to different choices of~$C$). We will specify such a choice as
needed.

The series $\varphi_\ell(g)$ are typically Gevrey-1, i.e., their coefficients have the factorial growth
\be
a_{n, \ell} \sim n!, \qquad n \gg 1,
\ee
at fixed $\ell$. This means that $\varphi_\ell(g)$ have zero radius of convergence. One way to make sense of these series is
to use generalized or lateral Borel resummations $s_\pm (\varphi)(g)$ (see, e.g.,~\cite{mmbook} for an~introductory
survey of Borel resummation techniques). We will assume that these resummations exist, for sufficiently small $g$.
The Borel resummed trans-series is then given by
\be
\label{br-transseries}
s_\pm (\Phi)= \sum_{\ell \ge 0} y^{\ell} s_\pm (\varphi_\ell)(g).
\ee
As we see, this object is itself an infinite sum over all exponentially small corrections.
Even if the lateral Borel resummations exist and one can make sense in~this way of the
factorially divergent series $\varphi_\ell (g)$, the sum over the infinite number of
non-perturbative sectors is potentially a new source of divergences.
In the case of nonlinear ODEs, this infinite sum is known to converge provided
${\rm Re}(g/A)$ is not too large~\cite{costin}, but in general one has to address this convergence
issue.

We can reorganize $\Phi$ in a different way, suggested by the work of
\cite{costin-costin} on non-linear ODEs: for each fixed power of $g$, $n$, we consider the series
\be
\label{CGdef}
\CG_n(y)= \sum_{\ell \ge 0} a_{n, \ell} y^\ell,
\ee
so that we can write, formally,
\be
\label{phi-y}
\Phi= \sum_{n \ge 0} g^n \CG_n (y).
\ee
We will call $\CG_n(y)$ the $n$-th partial series of the trans-series $\Phi$.
An important result of~\cite{costin-costin} is that, for non-linear ODEs, the partial series $\CG_n(y)$ are convergent, and
they have the same radius of convergence for all $n$. Generically, we expect
\be
a_{n, \ell} \sim \ell^{-b_n} R^{-\ell}, \qquad \ell \gg 1,
\label{an-bn-R}
\ee
for fixed $n$. The common radius of convergence is $R$, while the exponents $b_n$ depend in general on~$n$.

In many situations, like the one considered in~\cite{costin-costin},
the convergence property of the partial series is closely
related to the convergence of the sum over exponential corrections in~the Borel resummed trans-series (\ref{br-transseries}).
Heuristically, this can be argued as follows. If $g$ is sufficiently small, we can
approximate the lateral Borel resummations appearing in (\ref{br-transseries})
by uniform optimal truncations, i.e., we can truncate all factorially divergent series
$\varphi_\ell(g)$ in (\ref{trans-not}) at the same~${n=N^\star}$.
This is because all these series have the same rate of divergence. The
resulting approximation is nothing but the sum of the first $N^\star$ functions $\CG_n(y)$, which is convergent if the
partial series converge.
Note that this analysis cannot rule out the existence of higher trans-monomials \big(e.g.,~\smash{${\rm e}^{-{\rm e}^{1/g}}$}\big) which could be undetected by asymptotic behaviour. It is, in principle, also possible that the sum over such terms, or over corrections to such terms, could bring about other divergences in~the total trans-series.

Let us note that the radius $R$ signals the occurrence of a common singularity in~the functions~$\CG_n(y)$. One can
ask what is the relation between the location of the singularity in~the $\CG_n$s, and the location of the singularities of the
exact function represented by the resummed trans-series $\Phi$. As explained in~\cite{costin-costin}, the singularity in~the $\CG_n$'s gives the first term in an asymptotic approximation to the singularities of the exact function. In the case of nonlinear ODEs, one also
has that, for a fixed $|y|<R$ in~the common domain of convergence, the functions $\CG_n(y)$ grow
factorially~\cite{costin-costin},
\be
\label{new-fact}
\CG_n(y) \sim n!, \qquad n\gg 1.
\ee
Therefore, the expansion (\ref{phi-y}) provides an asymptotic representation of the
Borel-resummed trans-series for small $g$ and $|y|<R$ which is very different from the asymptotic representation
provided by $\varphi_0(g)$. We believe that the behavior (\ref{new-fact}) also occurs in~the QFT trans-series which satisfy the
convergence property.

In the following sections, we will consider interesting examples in QFT where the convergence property of the trans-series
holds.

\section{The large \texorpdfstring{$\boldsymbol{N}$}{N} sigma model, revisited}\label{sec-sigma}

One of the most explicit results for a trans-series expansion in QFT was obtained in~\cite{beneke-braun} for the
self-energy of the ${\rm O}(N)$ non-linear sigma model, at the first non-trivial order in~the $1/N$ expansion. Let us briefly review
the result and analyze the properties of the trans-series.

The basic field of the ${\rm O}(N)$ non-linear sigma model is a scalar
field $\boldsymbol{\sigma}\colon \IR^2 \rightarrow \IR^N$ satisfying the constraint
\be
\boldsymbol{\sigma}^2=1.
\ee
The Lagrangian density is
\be
\CL={1\over 2 g_0^2} \partial_\mu \boldsymbol{\sigma} \cdot \partial^\mu {\boldsymbol{\sigma}},
\ee
where $g_0$ is the bare coupling constant. As it is well-known, this model can be solved order by order in~the $1/N$ expansion.
In particular, one can find an explicit formula for the self-energy~$\Sigma(p)$ at NLO in $1/N$. This was calculated in~\cite{cr-dr} by using a convenient sharp momentum (SM) cutoff regularization, and one finds
\be
\Sigma(p)=p^2+ m^2 +{m^2 \over N} S\biggl( {p^2 \over m^2} \biggr) + \CO\biggl({1\over N^2} \biggr),
\ee
Here, $m$ is the mass gap, and $S(x)$ is given by
 \be
 \label{Sx}
 S(x)= \int_0^\infty \rd y \biggl( \log\biggl( {\xi+ 1 \over \xi-1} \biggr)\biggr)^{-1} \biggl[ {y \xi \over {\sqrt{(1+ y +x)^2-4 x y}}} -1+
 {x+ 1 \over 2} \biggl( {1\over \xi}-1\biggr) \biggr].
 \ee
 In this expression, we have denoted
 \be
 \xi ={\sqrt{1+ {4 \over y}}}.
 \ee
Since $m$ is proportional to the dynamically generated scale, it is natural to introduce a renormalized
coupling at the scale $p$, $g(p)$, by
\be
{1\over x}={ m^2 \over p^2}= \re^{-1/g(p)}.
\ee
We expect that $S(x)$, for small $x$, is given by a perturbative series in $g(p)$, together
with non-perturbative, exponentially small corrections. The perturbative expansion was obtained in~\cite{cr-dr}, while the full trans-series was obtained in~\cite{beneke-braun}
in a slightly different renormalization scheme. We~will now present the final result of this remarkable calculation.\footnote{Some details are clarified and worked out in~\cite{sss}, which extends the result of~\cite{beneke-braun} to the supersymmetric non-linear sigma model. Similar results for the Gross--Neveu model are discussed in~\cite{kneur}.}

The ingredients of the trans-series are contained in~three sequences of 
functions of an auxiliary variable $t$,
which can be regarded in fact as a Borel transform variable. The first sequence is denoted by $F_\ell(t)$, $\ell \ge 0$,
and given explicitly by
\be
F_\ell(t)= \sum_{j=0}^\ell 2\bigl(2t ^2 -t-j\bigr){\Gamma(2t+2j-1) \over \Gamma(2t+j+1)}{1 \over j!}\biggl[ {\Gamma(\ell+t-1) \over \Gamma(t+j-1) \Gamma(\ell-j+1)} \biggr]^2.
\ee
It can be easily checked that the $F_\ell(t)$ are polynomials in $t$ of degree $2\ell$. One finds, for example
\be
F_0(t)=1, \qquad F_1 (t)= t^2-1, \qquad F_2(t)= \frac{1}{4} \bigl(t^4+6 t^3+t^2-4 t-8\bigr).
\ee
The second sequence of functions is denoted by $G_\ell(t)$, with $\ell\ge 0$, and given explicitly by
\begin{align}
G_\ell(t)={}&{-}F_\ell(t) (\psi(t+\ell-1)+\psi(2-t-\ell))\\
&{}{+} \sum_{j=0}^\ell 4\bigl(2t ^2 -t-j\bigr)\psi(\ell+1-j) {\Gamma(2t+2j-1) \over \Gamma(2t+j+1)}{1 \over j!}\biggl[ {\Gamma(\ell+t-1) \over \Gamma(t+j-1) \Gamma(\ell-j+1)} \biggr]^2.\nonumber
\end{align}
Here, $\psi(t)$ is the digamma function. The functions $G_\ell(t)$ are meromorphic and have poles at all non-negative integers and some negative integers. One finds, for example,
\begin{align}
&G_0(t)= -2 \gamma_E - \psi(-1+t) -\psi(2-t),\nonumber\\
& G_1(t)=(t-1) (-2 ((\gamma_E -1) t+\gamma_E +1)-(t+1) \psi (1-t)-(t+1) \psi (t) ),
 \end{align}
where $\gamma_E$ is the Euler--Mascheroni constant. The third, and final sequence of functions, is
denoted by $H_\ell(t)$. One has
\begin{gather}
H_0(t)= {1\over t} -{1 \over t+1}, \qquad
H_1(t)=-\frac{1}{t}+\frac{2}{t+1}-\frac{1}{t-1}, \\
H_\ell(t)= (-1)^{\ell-1} \sum_{j=0}^{\ell-1}\biggl( {(\ell-2)! \over (\ell-j-2)! j!} \biggr)^2 2\bigl(2 t^2+j-\ell\bigr) (2\ell-2j-2)! \prod_{a=j-\ell}^{\ell-j} (t+a)^{-1}, \quad \ell \ge 2.\nonumber
\end{gather}
$H_\ell(t)$ is a rational function of $t$ with poles at $t=0, \pm 1, \dots, \pm \ell$. One has in addition the following
property:
\be
\label{resHresG}
{\rm Res}_{t=k} H_{\ell+k}(t)= {\rm Res}_{t=k} G_{\ell}(t), \qquad k \in \IZ_{\ge 0},
\ee
for all $\ell\ge 0$. We note that the functions $H_{0,1}(t)$ differ from the ones appearing in~\cite{beneke-braun}, due to a~difference
choice of renormalization scheme.

We have now all ingredients to construct the formal trans-series. Let us first define
\be
r_{\ell,k} = {\rm Res}_{t=k} \, H_\ell(t)
\ee
as well as the function
\be
\widehat G_\ell(t)= G_\ell(t) - \frac{r_{\ell,0}}{t},
\ee
which is regular at the origin, due to (\ref{resHresG}). We now define the formal series
\be
\label{sigma-trans}
\varphi_\ell(g)= F_\ell (0)+ \sum_{n\ge 1} \bigl( F_\ell^{(n)}(0) + \widehat G_\ell^{(n-1)}(0) \bigr) g^{n}, \qquad \ell \ge 0.
 \ee
 We will also need the integrals
\begin{equation}
c_\ell = -{\rm{P}} \int_0^\infty \biggl( H_\ell(t) -r_{\ell,0} \frac{ \re^{-t}}{t} \biggr) \rd t.
\end{equation}
Let us now consider the trans-series
\be
\label{beneke-ts}
\Phi_\pm =\sum_{\ell \ge 0} (-1)^\ell \re^{-\ell/g} \Biggl( r_{\ell,0}\log(g) + c_\ell \pm \ri\pi \sum_{k=1}^\ell r_{\ell,k} + \varphi_\ell (g) \Biggr).
\ee
This is slightly more general than the trans-series considered in (\ref{phits}) since it involves a logarithmic term. The
main result of~\cite{beneke-braun} is that the non-trivial integral (\ref{Sx}) appearing in~the self-energy can be obtained
as a lateral resummation of $\Phi_\pm$,
\begin{equation}\label{restrans}
{1\over x} S(x) =s_\pm (\Phi_\pm)= \sum_{\ell \ge 0} (-1)^\ell \re^{-\ell/g} \Biggl( r_{\ell,0}\log(g) + c_\ell \pm \ri\pi \sum_{k=1}^\ell r_{\ell,k} + s_\pm (\varphi_\ell)(g) \Biggr).
\end{equation}
Note that the constant, imaginary term of the trans-series is ambiguous and
depends on the choice of lateral resummation,
as expected since the work of David~\cite{david2,david3}. As noted in~\cite{beneke-braun},
the property (\ref{resHresG}) guarantees the cancellation of non-perturbative ambiguities, i.e.,
it guarantees that the two lateral resummations in (\ref{restrans}), with the correlated choice of sign in~the
imaginary piece, give the same result.

From the above results one obtains, e.g., the asymptotic behavior for the self-energy, which is given by the conventional perturbative series
\be
\frac{1}{x}S(x) \sim \log(g) + 1 - \gamma_E -2 g + \sum_{k \ge 1 }k! \bigl( \bigl(1+(-1)^k\bigr)\zeta(k+1) - 2 \bigr) g^{k+1}.
\ee
This was first found in~\cite{cr-dr} (the constant term of this series differs from the result quoted in~\cite{beneke-braun},
since our functions $H_{0,1}(t)$ are different from theirs).

We can now address the main question raised in~this paper. We have a family of formal series~(\ref{sigma-trans}) and, in~the notation of (\ref{trans-not}), their coefficients are given by
\be
a_{0, \ell}= F_\ell (0), \qquad a_{n,\ell}= F_\ell^{(n)}(0) + G_\ell^{(n-1)}(0), \quad n \ge 1.
\ee
We want to determine the behavior of these coefficients for large $\ell$ and fixed $n$. We find, numerically,
\be
\label{anl-sigma}
a_{n, \ell} \sim C_n {9^\ell \over \ell^2}, \qquad \ell \gg 1,
\ee
where $C_n$ is an $n$-dependent constant. We have verified this behavior for
many values of $n\le 20$, and we conjecture that this asymptotic behavior holds for
all $n \ge 0$. This implies in particular that the series $\CG_n(y)$ are convergent and
they have the same radius of convergence for all $n$, namely, $R=1/9$. We have also found
numerically that $C_n \sim n!$, as one would
expect from (\ref{new-fact}). Let us also note that the coefficients $c_{\ell}$, $r_{\ell, 0}$
appearing in (\ref{beneke-ts}), as well as the sum
\be
\sum_{k=1}^\ell r_{\ell, k}
\ee
have the same asymptotic behavior for large $\ell$ as the coefficients $a_{n, \ell}$, namely $9^\ell/ \ell^2$.
In particular, we find a singularity of
the partial series at
\be
\label{sing-sigma}
{m^2 \over p^2}= -{1 \over 9}.
\ee

\begin{figure}[t]
\centering
\includegraphics[width=0.4\textwidth]{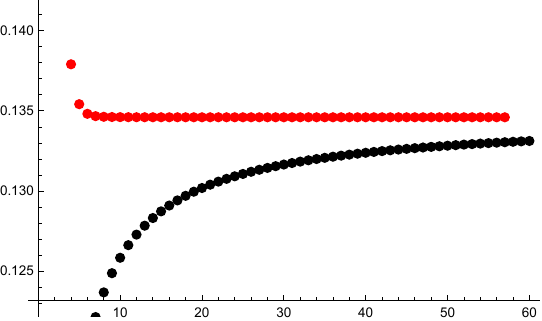} \qquad \qquad
\includegraphics[width=0.4\textwidth]{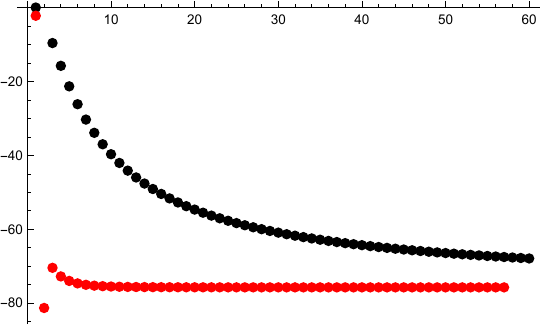}
\caption{The sequences (\ref{quot-seq}) for $n=1$ (left) and $n=6$ (right), represented by black dots, together with their Richardson
transforms of order $3$, represented by red dots.}
\label{vert-series}
\end{figure}

To illustrate the behavior (\ref{anl-sigma}), we plot in Figure~\ref{vert-series} the sequence
\be
\label{quot-seq}
s^{(n)}_{\ell}= {a_{n,\ell } \over 9^\ell \ell^{-2}}
\ee
for $\ell=1,\dots, 60$, and $n=1,6$, together with their Richardson transform of order $3$. It~is~manifest that they converge rapidly to an $n$-dependent constant.

As we explained in Section~\ref{tss-sec}, we expect that the convergence property of
the partial series implies the convergence of the
full trans-series. In this example, this is realized in a simple way, and the
common radius of convergence that we have found for the series $\CG_n(y)$ is actually the same one that was found
in~\cite{beneke-braun} for the full resummed trans-series. However, the two radius of convergence
do not need to match, in general. The present trans-series is peculiar in~this regard, related to the
fact that the parameters $b_n$ of~\eqref{an-bn-R} are in~this case the same for all $n$. In the trans-series of the free energy discussed in Section~\ref{sec-integrable}, the $b_n$ does change with $n$, and the radius of convergence extracted from the partial series can only provide an approximation for the radius of the full trans-series.

\begin{figure}[t]
\centering
\includegraphics[width=.7\textwidth]{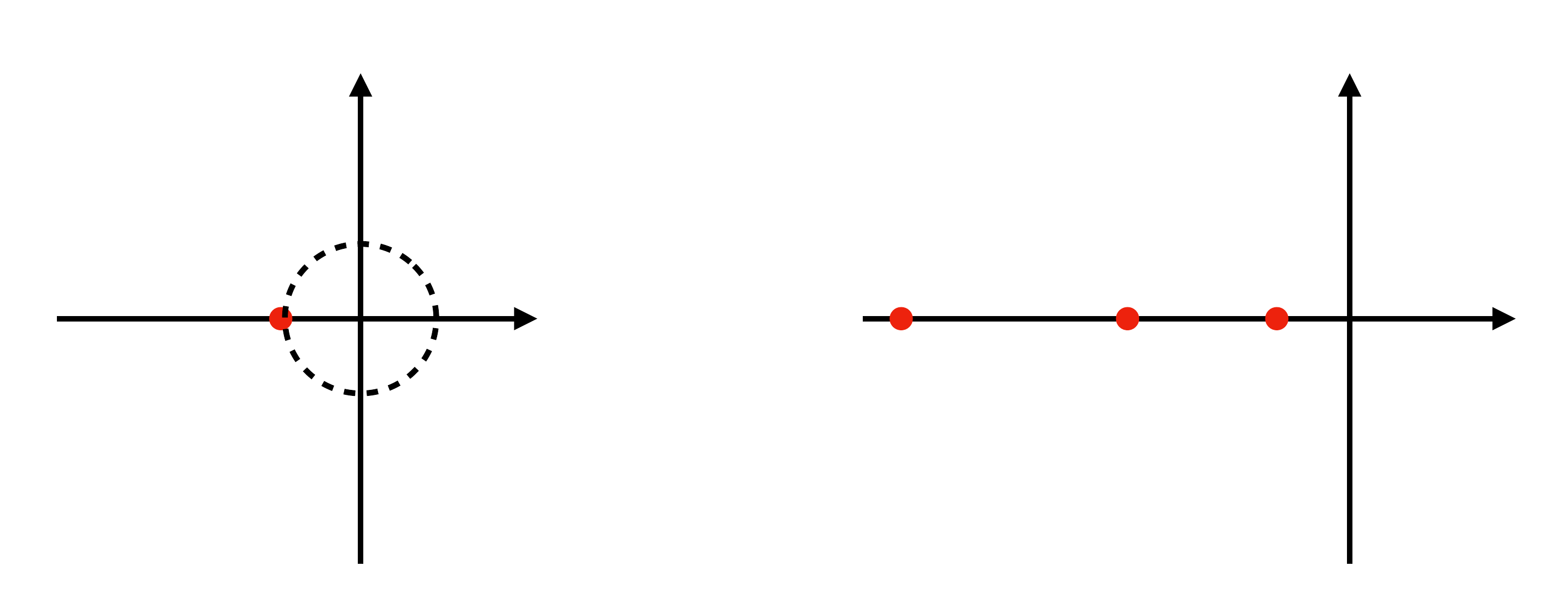}
\caption{The singularity structure of the large $N$ two point function
is shown on the left, in~the $p^2/m^2$ plane. There is a single branch cut
singularity at $-9$, as it follows from (\ref{sing-sigma}). At finite $N$, it has
been argued in~\cite{beneke-braun} that there will be
many singularities along the negative real axis, at positions $-(2k+1)^2$,
$k \in \IZ_{>0}$, as shown on the figure on the right.}\label{fig-sing1}
\end{figure}

We have verified the convergence property of partial series for the trans-series representing
the large $N$ two-point function, but we expect this property to fail at finite $N$. The reason is the following.
It is easy to check from the
explicit expression for $S(x)$ that the two-point function at large $N$ has a single singularity at
precisely the point (\ref{sing-sigma}), which is also the common location of the singularity for the
series $\CG_n(y)$, see Figure~\ref{fig-sing1}. Physically, this represents the threshold for production of
three particles of mass $m$, as noted in~\cite{beneke-braun}, and it is
a consequence of working at large $N$. At finite $N$ there will be infinitely many multi-particle
thresholds, and we expect singularities at $p^2/m^2= -(2k+1)^2$, $k \in \IZ_{>0}$, as show on the right at Figure~\ref{fig-sing1}.
As pointed out in~\cite{shifman, shifman-hadrons}, this analytic structure can not be
obtained from a convergent trans-series, and in particular from a trans-series with the
convergent property of partial series.
Heuristically, we~expect that a trans-series with the latter property will lead to singularities
in a bounded set of the $p^2/m^2$ plane, the farthest singularity being approximated by the common singularity of the partial series.
This is the case for the two-point function at large $N$, but not at finite $N$.

As an additional observation, let us mention that the trans-series (\ref{beneke-ts}) is a clear example
in which the strong version of the resurgence program discussed in~\cite{dmss} does {\it not} hold.
According to the strong version, all the
non-perturbative corrections appearing
in~the trans-series should be obtained from the analysis of the
Borel singularities of the perturbative series, perhaps up to
overall multiplicative constants. However, it is easy to see that the first
Borel singularity of~$\varphi_0(g)$ leads to a trivial trans-series, proportional to
$\re^{-1/g}$, and can not in any way reconstruct the full $\varphi_1(g)$. This is similar to the examples
found in~\cite{dmss}, and it might be due to the fact that the trans-series
(\ref{beneke-ts}) is obtained at a fixed order in~the
large $N$ expansion.

\section{Complex Chern--Simons theory}\label{sec-cs}

Complex Chern--Simons (CS) theory is an ideal arena to study trans-series,
since as a QFT it is essentially solvable, and at the same time
it leads to highly non-trivial perturbative series. The mathematical program of applying resurgence in
complex CS theory was pioneered in~\cite{garoufalidis}, and applied in, e.g.,~\cite{garo-costin,gmp}.

In this section we will illustrate our considerations with the resurgent
results for partition functions on the complements of knots, presented in~\cite{ggm2, ggm1}. This
observable admits a non-perturbative definition through the so-called state integral invariant~\cite{DGLZ, ak, hikami},
which can be calculated by a finite-dimensional integral. The coupling constant will be denoted by $\tau$ and is a
complex variable. The partition function depends on an additional complex variable, the~holonomy
around the knot, but we will set it to zero for simplicity. In this case, the
partition function is known to be an analytic function of $\tau$ on $\IC \backslash (-\infty, 0]$.

For concreteness, we will consider the complement of the famous figure-eight knot. Its state-integral
invariant is explicitly known, and we would like to express it as a trans-series. The~building blocks of the
trans-series are the perturbative expansions around non-trivial saddle points or instantons in
complex CS theory. In this case, there are two of them, corresponding to the so-called
geometric connection and its complex conjugate. We will denote the perturbative series
attached to these saddle-points by $\varphi_{g,c}(\tau)$, where $g (c)$ refers to the geometric
(respectively, conjugate) connection. They can be computed explicitly up to the very high order, and their very first terms read as
\be
\varphi_g(\tau)= \sum_{k \ge 0} a_k \tau^k= 3^{-1/4} \biggl( 1+ {11 \pi \over {\sqrt{3}}\cdot 36} \tau+ {697 \pi^2 \over 7776} \tau^2+ \cdots \biggr), \qquad
\varphi_c(\tau)= \varphi_g(-\tau).
\ee
The partition function (or Andersen--Kashaev invariant) will be denoted by $\CI (\tau)$. As shown in~\cite{ggm1}, it
can be expressed as the lateral Borel resummation of a trans-series involving the above perturbative series.
Not surprisingly, this expression depends on the region of the $\tau$ plane which we consider. When $\tau$ is slightly above the positive real axis, one simply finds
\be
\label{I1}
\CI(\tau) = -2 \ri \re^{-{V \over 2 \pi \tau}} s(\varphi_c)(\tau),
\ee
where $V=2.029883\dots$ is the volume of the figure-eight knot, and it is the classical action of the conjugate connection in complex
CS theory. In~(\ref{I1}), the resummation is along the direction set by the argument of $\tau$. However, as
we move to the region slightly above the negative real axis, we find the expression
\be
\label{I2}
\CI(\tau)=-2 \ri \re^{-{V \over 2 \pi \tau}} s(\varphi_c)(\tau) M_1 (\tilde q)
+ 2 \re^{{V \over 2 \pi \tau}} s(\varphi_g)(\tau) M_2(\tilde q).
 \ee
Here,
\be
\tilde q= \re^{-{2 \pi \ri \over \tau}},
\ee
and the resummation is also along the direction set by the argument of $\tau$.
$M_{1,2}(\tilde q)$ are power series in $\tilde q$. Their very first terms read
\begin{align}
&M_1 (q)= 1 - 8 q -9 q^2 + 18 q^3+ 46 q^4+\CO\bigl(q^5\bigr),\nonumber\\
&M_2 (q)=-9 q - 3 q^2 + 39 q^3 + 69 q^4 + \CO\bigl(q^5\bigr),
\end{align}
and their exact expressions were conjectured in~\cite{ggm1}. We will not need them
here, but we note that they are convergent power series with radius of convergence $R=1$.

The result (\ref{I2}) has a very clear physical interpretation. The terms
$\tilde q^n \exp(\pm V/(2 \pi \tau))$ correspond to contributions of
complex instantons, whose actions are of the form
\be
\mp {V \over 2 \pi}+ 2 \pi \ri n, \qquad n \in \IZ_{>0}.
\ee
The additional term $2 \pi \ri n$ is due to the multivaluedness of the CS action. Therefore, the expression (\ref{I2}) can be
regarded as a multi-instanton expansion in which we sum over an infinite
number of exponentially small corrections (which in fact arise as Borel
singularities of the series~$\varphi_{g,c}(\tau)$, see~\cite{ggm1} for a detailed explanation). In terms of the formalism we
set in Section~\ref{tss-sec}, we have a slightly more general trans-series than (\ref{phits}). Let us denote
\be
\label{q-series}
M_1( q)= \sum_{n \ge 0} \mu_{1, n} q^n, \qquad M_2( q)= \sum_{n \ge 0} \mu_{2, n} q^n.
\ee
Then, we can write the underlying trans-series as
\be
\Phi(\tau)= -2 \ri \re^{-{V \over 2 \pi \tau}} \sum_{\ell \ge 0} \tilde q^\ell \varphi_{c, \ell} (\tau) + 2 \re^{{V \over 2 \pi \tau}}\sum_{\ell \ge 1} \tilde q^\ell \varphi_{g, \ell}(\tau),
\ee
where
\be
\varphi_{c, \ell} (\tau)= \mu_{1, \ell} \varphi_{c} (\tau), \qquad \varphi_{g, \ell} (\tau)=\mu_{2, \ell} \varphi_{g} (\tau).
\ee
Due to the presence of two different actions, we have two different functions $\CG_n(y)$ to consider, and they are given by
\begin{gather}\label{csGs}
\CG_n^1(y)= (-1)^n a_n \sum_{\ell \ge 0} \mu_{1, \ell} y^\ell= (-1)^n a_n M_1(y), \qquad \CG_n^2 (y)= a_n \sum_{\ell \ge 0} \mu_{2, \ell} y^\ell=a_n M_2(y),\!
\end{gather}
where in~this case $y= \tilde q$.
Therefore, it is clear that the convergence of the series $M_{1,2}(q)$ defined in (\ref{q-series})
guarantees the convergence of the partial series, and
it is obvious from (\ref{csGs}) that the radius of convergence of $\CG^i_n(y)$, $i=1,2$ are the same, for all $n$.

Although we have focused on a particular example, the resulting structure seems to be typical of many trans-series in complex CS theory:
the partial series are given by convergent $\tilde q$ series. As we have pointed out, this makes it possible to have a simple resurgent structure without further resummations.

\section{Integrable models}\label{sec-integrable}

\subsection{The free energy of integrable models}

Two-dimensional, integrable, asymptotically free quantum field theories are an interesting family of toy
models where one can obtain an explicit but highly non-trivial trans-series
for the free energy~\cite{abbh1, abbh2,bbhv, bbhv2,bbv,bbv2,dmss,fkw-letter,fkw, mmr,mmr-an, mmr-theta,mr-ren,s-thompson, volin}, and therefore one can hope that the~convergent property of the partial series can be
tested in detail.

\begin{figure}[t]
\centering
\includegraphics[width=.3\textwidth]{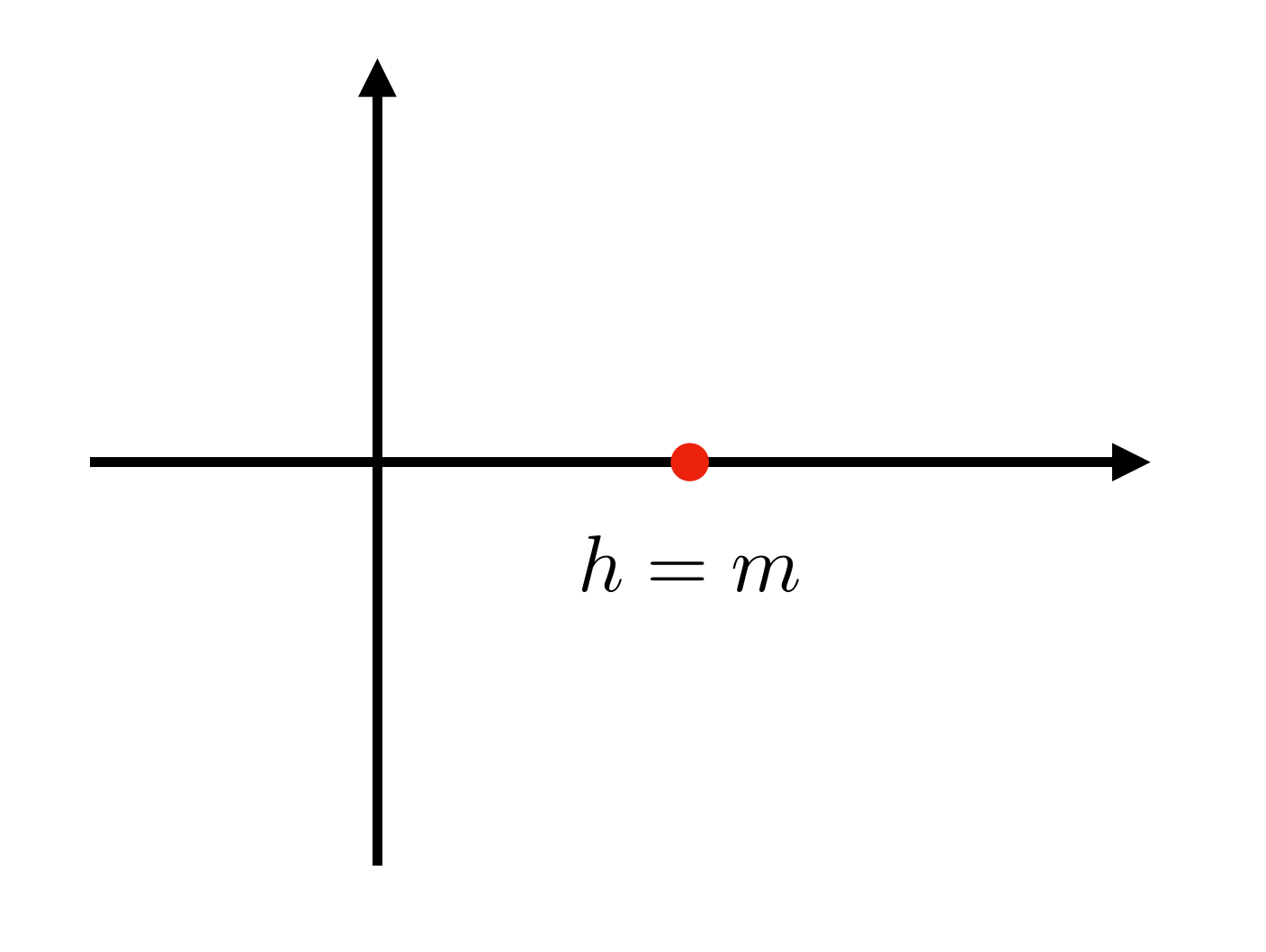}
\caption{The free energy $\CF(h)$ has a branch cut singularity at $h=m$.}\label{fig-singfh}
\end{figure}

In integrable quantum field theories, the free energy is a quantity that can be computed after we
introduce into the Hamiltonian $\mathsf{H}$ a chemical potential $h$ coupled to a conserved charge $\mathsf{Q}$,
\begin{equation}
\mathsf{H} \rightarrow \mathsf{H} - h \mathsf{Q}.
\label{eq_hamiltonian_mod}
\end{equation}
For each model, there can be different convenient choices of the charge $\mathsf{Q}$.
The ground state of the Hamiltonian~\eqref{eq_hamiltonian_mod} is a finite density state characterized by the free energy per unit volume
\begin{equation}
F(h) = -\lim_{V, \beta \rightarrow \infty} {1\over V\beta } \log {\Tr}\, \re^{-\beta (\mathsf{H}-h \mathsf{Q})},
\end{equation}
where $V$ is the total volume of space and $\beta$ is the total length of Euclidean time. For convenience, we will write our
equations in~terms of
\begin{equation}\label{subF}
\mathcal{F}(h) = F(h) - F(0).
\end{equation}
As found by Polyakov and Wiegmann~\cite{pw,wiegmann}, $\CF(h)$ can be computed by the Bethe ansatz. This amounts to
the following integral equation for a Fermi density $\epsilon(\theta)$:
\begin{equation}
\epsilon(\theta)-\int_{-B}^B K(\theta-\theta') \epsilon (\theta') \rd \theta' = h-m \cosh(\theta),\qquad \theta \in[-B,B].
\label{eq_epsilon_integral_equation}
\end{equation}
In this equation, $m$ is the mass of the particles charged under $\mQ$. The kernel of the integral equation $K(\theta)$ is given by
\begin{equation}
K(\theta) = \frac{1}{2\pi\ri} \frac{\rd}{\rd \theta} \log S(\theta),
\end{equation}
where $S(\theta)$ is the $S$-matrix appropriate for the scattering of the charged particles.
The endpoints~$\pm B$ of the Fermi density are fixed by the boundary condition
\begin{equation}
\epsilon(\pm B) = 0.\label{eq_boundary_condition_0}
\end{equation}
As we will see later, the boundary condition determines the parameter $B$ in~terms of $h/m$.
Once we find the solution $\epsilon(\theta)$ to~\eqref{eq_epsilon_integral_equation}, the free energy is given by
\begin{equation}
\mathcal{F}(h) = -\frac{m}{2\pi} \int_{-B}^B \epsilon(\theta) \cosh(\theta) \rd\theta.\label{eq_free_energy_integral}
\end{equation}
Note that we calculate $\CF(h)$ for $h>m$, such that the ground state has particles in it.
At the threshold value $h=m$, the one particle state has the same energy as the vacuum and we have $\CF(m)=0$.
In fact, $\CF(h)$ has a branch cut singularity as $h\rightarrow m^+$, as it can be verified
by explicit calculations, see Figure~\ref{fig-singfh}. For example, for the ${\rm O}(3)$ non-linear sigma model, one has~\cite{hmn}
\be
\CF(h) \sim (h-m)^{3/2}, \qquad h\rightarrow m^+.
\ee

While it might be easy to numerically extract the
free energy from~\eqref{eq_epsilon_integral_equation} and
\eqref{eq_free_energy_integral}, the key challenge is to study these
equations in~the weakly coupled regime $h \gg m$, which
corresponds to the limit $B \gg 1$, and to extract the complete
asymptotics, including exponentially small corrections. In
\cite{volin}, a powerful technique to extract perturbative
corrections was found, while in~\cite{mmr-an} an analytic method
to extract exponentially small corrections analytically, directly from the Bethe ansatz,
was developed. More recently,~\cite{bbv2} put together these
two developments and obtained explicit results for the full
trans-series. Since
we will only consider the very first partial series, we will follow
the method in~\cite{mmr-an}. It is possible that more systematic results might be
obtained by implementing the method of~\cite{bbv2}.

The first step is to reexpress~\eqref{eq_epsilon_integral_equation},
\eqref{eq_boundary_condition_0} and~\eqref{eq_free_energy_integral} in~the Fourier space of the
$\theta$ variable and use the Wiener--Hopf approach~\cite{fnw1,hmn, jap-w}.
We consider the Fourier transform of the kernel
\begin{equation}
\tilde{K}(\omega) = \int_\mathbb{R} \re^{\ri \omega \theta} K(\theta) \rd\theta
\end{equation}
and its Wiener--Hopf factorization
\begin{equation}
1-\tilde{K}(\omega) = \frac{1}{G_+(\omega) G_-(\omega)},
\end{equation}
where $G_\pm(\omega)$ is analytic in~the upper (respectively, lower)
complex half plane. The functions~$G_\pm(\omega)$ will depend on the
specific model, but we will only
consider the case in which~$K(\theta)$ is an even function, which implies that $G_-(\omega) = G_+(-\omega)$. It will be convenient to define
\begin{equation}
\sigma(\omega) = \frac{G_-(\omega)}{G_+(\omega)}.
\label{eq_sigma_def}
\end{equation}
We further introduce
\begin{equation}
g_+(\omega) = \ri h \frac{1-\re^{2\ri B\omega}}{\omega} +
 \frac{\ri m\re^B}{2}\biggl( \frac{\re^{2\ri B\omega}}{\omega-\ri} - \frac{1}{\omega+\ri} \biggr),
\label{eq_g}
\end{equation}
which is the Fourier transform of a function related to the
expression appearing on the right-hand side of the integral equation
\eqref{eq_epsilon_integral_equation} (for details, see
\cite{mmr-an}). Lastly, we define
\begin{equation}
\epsilon_\pm(\omega) = \re^{\pm\ri B \omega} \tilde{\epsilon}(\omega),
\end{equation}
where $\tilde{\epsilon}(\omega)$ is the Fourier
transform of the Fermi density $\epsilon(\theta)$.

As shown in~\cite{fnw1, hmn}, the integral equation~\eqref{eq_epsilon_integral_equation} can be
reformulated as the following equation:
\begin{equation}
Q(\omega) = \frac{1}{2 \pi \ri}
\int_\mathbb{R} \frac{G_-(\omega') g_+(\omega')}{\omega+ \omega'+ \ri 0} \rd \omega'
+ \frac{1}{2 \pi \ri} \int_\mathbb{R} \frac{\re^{2 \ri B\omega'} \sigma (\omega') Q(\omega')}{\omega+ \omega'+ \ri 0} \rd \omega',
\label{eq_Q_integral_equation}
\end{equation}
which has to be solved for $Q(\omega)$. The solution $Q(\omega)$ determines
$\epsilon_+(\omega)$ through the equation
\begin{equation}
\frac{\epsilon_+(\omega)}{G_+(\omega)} = \frac{1}{2 \pi \ri}
\int_\mathbb{R} \frac{G_-(\omega') g_+(\omega')}{\omega'-\omega- \ri 0} \rd \omega'+
\frac{1}{2 \pi \ri} \int_\mathbb{R} \frac{\re^{2 \ri B\omega'} \sigma (\omega')Q(\omega')}{\omega'- \omega- \ri 0} \rd \omega'.
\label{eq_epsilon_0}
\end{equation}
A relationship between $B$ and $h/m$ is determined by the boundary
condition~\eqref{eq_boundary_condition_0}, which in Fourier space takes the form
\begin{equation}
\lim_{\kappa \to +\infty} \kappa \epsilon_+(\ri \kappa)=0.
\label{eq_boundary_condition}
\end{equation}
Finally, from~\eqref{eq_free_energy_integral}, the free energy can be written as
\begin{equation}
\mathcal{F}(h)= -{1\over 2 \pi} m \re^B \epsilon_+(\ri).
\label{eq_free_energy_fourier}
\end{equation}

\subsection{Bosonic models}

\begin{figure}
\centering
\includegraphics[scale=0.9]{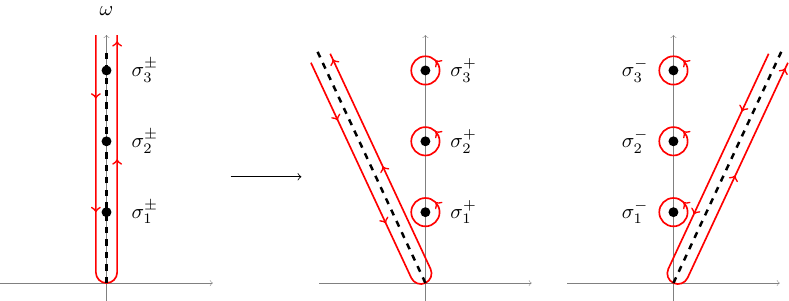}
\caption{Generic singularity structure of $\sigma(\omega)$. The second integral in~\eqref{eq_Q_integral_equation} or~\eqref{eq_epsilon_0} is integrated over~$\mathbb{R}$, but we deform the contour upwards, picking the discontinuity of $\sigma(\omega)$ and its residues. The residues will depend on the branch choice for $\sigma(\omega)$, as illustrated by tilting the branch cut with a small angle.}
\label{fig_residue_ambiguity}
\end{figure}

Using the techniques described in~\cite{mmr-an}, one can extract the trans-series of the free energy from the equations above.
In this section we will concentrate on the bosonic models, which include the ${\rm O}(N)$ non-linear sigma model
and the ${\rm SU}(N)$ principal chiral field (PCF). The structure of the trans-series is generically given by
\begin{equation}
\Phi_\pm(\tilde{\alpha}) = \sum_{\ell\geq 0} y^\ell \varphi_\ell(\tilde{\alpha}) + \mathcal{C}_{\text{IR}}^\pm \tilde{\alpha}^{2-2a} \re^{-2/\tilde{\alpha}},
\label{eq_free_energy_transseries}
\end{equation}
where $\tilde{\alpha}$ is a conveniently defined coupling (see~\eqref{eq_alpha_tilde_def} below) and $y = \tilde\alpha^{-2a\xi_1} \re^{-2\xi_1/\tilde{\alpha}}$ (see below for the definition of $\xi_1$) is the appropriate exponentially small term in~the present example. The~second term in~the right-hand side
is an IR renormalon term and it is not relevant for the analysis of the trans-series convergence. The free energy can be recovered from Borel resummation of~\eqref{eq_free_energy_transseries} as
\begin{equation}
\mathcal{F}(h) = -\frac{k^2 h^2}{4 \tilde{\alpha}} s_\pm \bigl( \Phi_\pm \bigr)(\tilde{\alpha}) = -\frac{k^2 h^2}{4 \tilde{\alpha}}\biggl\{ \sum_{\ell\ge 0} y^\ell s_\pm (\varphi_\ell) (\tilde{\alpha}) + \mathcal{C}_{\text{IR}}^\pm \tilde{\alpha}^{2-2a} \re^{-2/\tilde{\alpha}} \biggr\}.
\end{equation}
The coupling $\tilde{\alpha}$ that we use for the trans-series expansion is defined as
\begin{equation}
\frac{1}{\tilde{\alpha}} + \biggl( a - \frac{1}{2} \biggr) \log \tilde{\alpha} = \log \frac{h}{\Lambda},
\label{eq_alpha_tilde_def}
\end{equation}
where $\Lambda$ is the dynamically generated scale in~the $\overline{\text{MS}}$ scheme. This is a convenient choice of coupling that makes possible the contact with standard perturbative results, when one computes the free energy directly from the Lagrangian. The parameters $k$ and $a$ can be extracted from the behavior of $G_+(\omega)$ near the origin,
\begin{equation}
G_+(\ri\xi) = \frac{k}{\sqrt{\xi}} \bigl( 1 - a\xi \log\xi - b\xi + \mathcal{O}\bigl(\xi^2\bigr) \bigr), \qquad \xi >0.
\label{eq_G_series}
\end{equation}
The structure of~\eqref{eq_G_series} is valid for generic bosonic models, but it does
not apply to fermionic models and, for this reason, we have to treat them separately in
Section~\ref{sec_gross_neveu}. Further, we denote by $\xi_\ell$ the singularities of $\sigma(\ri\xi)$
at $\xi>0$, which we assume to be equally spaced:\footnote{This holds in~the models we study here, but it is not a generic property of bosonic models.} $\xi_\ell = \xi_1 \ell$. This determines the set
of exponentially small terms that we find in~the trans-series~\eqref{eq_free_energy_transseries} of the free energy. Lastly, as we will see in~the explicit computation further below, the coefficients determining each formal power series $\varphi_\ell(\tilde{\alpha})$ can be expressed as multinomials in~the residues
\begin{equation}
\ri\sigma^\pm_\ell = \Res_{\xi=\xi_\ell\mp \ri 0} \sigma(\ri\xi),
\end{equation}
which have an imaginary ambiguity (denoted by the two possible sign choices) due to the presence of a branch cut along $\xi>0$. Thus, the coefficients of $\varphi_\ell(\tilde{\alpha})$ are also themselves ambiguous. See Figure~\ref{fig_residue_ambiguity} for the singularity structure of $\sigma(\omega)$ and an illustration of how the residues depend on the branch choice.

\subsubsection{Trans-series computation}
\label{sec_bosonic_models}
In this subsection, we will compute the formal power series $\varphi_\ell(\tilde\alpha)$ appearing in~the trans-series~\eqref{eq_free_energy_transseries}. Our computation is limited to the leading and subleading corrections in $\tilde{\alpha}$, but to arbitrary order~$\ell$ in~the exponential corrections. This will be essential in order to extract the large order behavior of the coefficients that appear in~the partial series $\mathcal{G}_n(y)$ of~\eqref{phi-y}, for $n=0, 1$.

The computation is split in~two main steps. First, we determine the trans-series of the free energy $\mathcal{F}(h)$ in~the small parameters $1/B$ and $\re^{-2B\xi_1}$, by working out the equations~\eqref{eq_Q_integral_equation} and~\eqref{eq_epsilon_0} and the boundary condition~\eqref{eq_boundary_condition}. In the second step, we use~\eqref{eq_alpha_tilde_def} to determine~$B$ as a trans-series in $\tilde{\alpha}$ and \smash{$y = \tilde\alpha^{-2a\xi_1} \re^{-2\xi_1/\tilde{\alpha}}$}. With this later result, we will be able to write the trans-series we obtained in~the first step as a trans-series in~the coupling $\tilde\alpha$.

From~\eqref{eq_Q_integral_equation}, we have to compute $Q(\ri\xi)$ at $\xi=\xi_m$ as a trans-series in $1/B$, which we will need later for $\epsilon_+(\ri\kappa)/G_+(\ri\kappa)$ and the free energy. After evaluating~\eqref{eq_Q_integral_equation} at $\omega=\ri\xi_m$, we obtain~the expression
\begin{equation}
Q(\ri\xi_m)= \frac{1}{2 \pi \ri} \int_\mathbb{R} \frac{G_-(\omega) g_+(\omega)}{\omega+ \ri\xi_m} \rd \omega + \frac{1}{2 \pi \ri} \int_\mathbb{R} \frac{\re^{2 \ri B\omega} \sigma (\omega) Q(\omega)}{\omega+ \ri\xi_m} \rd \omega.
\label{eq_Qm_integral_equation}
\end{equation}
Let us compute the first integral in~\eqref{eq_Qm_integral_equation}. After splitting $g_+(\omega)$ into terms proportional to $h$ and $m$, as in~\eqref{eq_g}, we obtain for the terms multiplied by $h$
\begin{gather}
\ri h \frac{1}{2\pi\ri} \int_\mathbb{R} \frac{G_-(\omega)}{\omega + \ri\xi_m} \frac{1-\re^{2\ri B\omega}}{\omega} \rd \omega \nonumber\\
\qquad{}= -\frac{kh (2B)^{1/2}}{\pi \xi_m} \biggl[ 2 \sqrt{\pi} - \frac{\log(2B)}{B}a I_0+ \frac{1}{B}\biggl( a I_1 + b I _0 - \frac{I_0}{\xi_m} \biggr) \biggr]
+ h \frac{G_+(\ri\xi_m)}{\xi_m} \nonumber\\
\qquad\quad{}+ h \sum_{n\ge 1} \frac{\ri \sigma^\pm_n G_+(\ri\xi_n)}{(\xi_n+\xi_m)\xi_n} \re^{-2B\xi_n} + \mathcal{O}\bigl(B^{-1}\bigr),
\label{eq_h_part_2}
\end{gather}
where we have defined the integrals
\begin{align}
&I_0= -\frac{1}{2}\int_0^\infty \frac{\re^{-y}}{\sqrt{y}} \rd y = -\frac{\sqrt{\pi}}{2},\\
&I_1= -\frac{1}{2}\int_0^\infty \frac{\re^{-y}}{\sqrt{y}}\log(y) \rd y = \frac{\sqrt{\pi}}{2}(\gamma_E+\log(4)).
\end{align}
To obtain~\eqref{eq_h_part_2}, we deform the contour downwards for the integral with $1/\omega$, and upwards for the integral with $\re^{2\ri B \omega}/\omega$. The downwards contour picks the pole at $\omega = -\ri\xi_m$, which gives the term proportional to $G_+(\ri\xi_m)$ in~the second line. The upwards contour picks the discontinuity of~$G_-(\omega)$, which yields the first term in square brackets, and also picks the singularities of~$G_-(\omega)$, which yields the sum over the exponentials in~the second line.

For the terms in $g_+(\omega)$ with $m$, we find
\begin{gather}
\frac{\ri m\re^B}{2} \frac{1}{2\pi\ri} \int_\mathbb{R} \frac{G_-(\omega)}{\omega + \ri\xi_m} \biggl( \frac{\re^{2\ri B\omega}}{\omega-\ri} - \frac{1}{\omega+\ri} \biggr) \rd \omega\label{eq_m_part_2}\\
= -\frac{m\re^B}{2}\biggl[ \frac{G_+(\ri\xi_m)-G_+(\ri)}{\xi_m-1}+\! \sum_{n\ge 1} \frac{\ri\sigma^\pm_n G_+(\ri\xi_n)}{(\xi_n + \xi_m)(\xi_n -1)}\re^{-2B\xi_n} - \frac{k}{\sqrt{\pi} \xi_m}(2B)^{-1/2} \biggr]\! + \mathcal{O}\bigl( B^{-1} \bigr).
\nonumber
\end{gather}
In this case, the first term arises from the two poles at $\omega = -\ri$ and $-\ri\xi_m$, when deforming the contour downwards for the term $-1/(\omega+\ri)$. The sum over the exponentials arises from the term~$\re^{2\ri B\omega}/(\omega-\ri)$, when deforming the contour upwards and picking all the poles of $G_-(\omega)$. The last term arises from the integral over the discontinuity of $G_-(\omega)$.

Next, we consider the second integral in~\eqref{eq_Qm_integral_equation}, for which we deform the contour upwards, picking both the discontinuity and the poles of $\sigma(\ri\xi)$,
\begin{gather}
\frac{1}{2 \pi \ri} \int_\mathbb{R} \frac{\re^{2 \ri B\omega} \sigma (\omega) Q(\omega)}{\omega+ \ri\xi_m} \rd \omega \nonumber\\
\qquad{}= \frac{1}{2\pi\ri} \int_0^\infty \frac{\re^{-2B\xi'}\delta\sigma(\ri\xi')Q(\ri\xi')}{\xi' + \xi_m} \rd\xi'
- \sum_{n\ge 1} \frac{\ri\sigma^\pm_n Q(\ri\xi_n)}{\xi_n + \xi_m} \re^{-2B\xi_n},
\label{eq_sigma_part}
\end{gather}
where $\delta \sigma(\ri\xi)$ denotes the discontinuity of $\sigma(\ri\xi)$. The above equation, which we need to extract~$Q(\ri\xi_m)$, involves the quantities $Q(\ri\xi_n)$, thus we will need to recursively solve for these quantities at each exponential order. Another complication is the integral over $Q(\ri\xi')$ in~\eqref{eq_sigma_part}, for which we need to determine $Q(\ri\xi)$ at small $\xi$. We propose the following trans-series ansatz:
\begin{equation}
Q(\ri x /2B) = \sum_{n\ge 0} Q_{n} (x) \re^{-2B\xi_n},
\end{equation}
where each $Q_{n}(x)$ is a function of $x$ that we expand in powers of $1/B$. These functions will be
solutions to integral equations that involve what we called in~\cite{mmr-an} the Airy operator:%
\begin{equation}
(\mathsf{K}f)(x) = \int_0^\infty \frac{\re^{-y}}{x+y} f(y)\mathrm{d}y.
\end{equation}
To obtain~these integral equations, we need to write~\eqref{eq_Q_integral_equation} at $\omega=\ri x/2B$:
\begin{equation}
Q(\ri x/2B)= \frac{1}{2 \pi \ri} \int_\mathbb{R} \frac{G_-(\omega) g_+(\omega)}{\omega+ \ri x/2B} \rd \omega + \frac{1}{2 \pi \ri} \int_\mathbb{R} \frac{\re^{2 \ri B\omega} \sigma (\omega) Q(\omega)}{\omega+ \ri x /2B} \rd \omega.
\label{eq_QB_integral_equation}
\end{equation}
Doing a similar computation to~\eqref{eq_h_part_2}, we find
\begin{gather}
\ri h \frac{1}{2\pi\ri} \int_\mathbb{R} \frac{G_-(\omega)}{\omega + \ri x/2B} \frac{1-\re^{2\ri B\omega}}{\omega} \rd \omega\nonumber\\
\qquad{}= -kh(2B)^{1/2} \biggl[ 2B \frac{\mathsf{K}}{\pi}\frac{\re^x - 1}{x^{3/2}}+ \biggl( 1 - \frac{\mathsf{K}}{\pi} \biggr) \frac{-a\log(2B) + a \log(x) + b}{x^{1/2}} \biggr]+ \mathcal{O}\bigl(B^0\bigr).
\end{gather}
The details of this computation were carried out in~\cite[Appendix A]{mmr-an}. In principle, in our result, we should also incorporate exponential corrections arising from the poles of $G_-(\omega)$ as we deform the contour upwards for the term $\re^{2\ri B \omega}$. However, these are of order $B^0$ and we can ignore them at the order we are working at.

By carrying out manipulation analogous to~\eqref{eq_m_part_2}, we find for the terms proportional to $m$
\begin{gather}
\frac{\ri m\re^B}{2} \frac{1}{2\pi\ri} \int_\mathbb{R} \frac{G_-(\omega)}{\omega + \ri x/2B} \biggl( \frac{\re^{2\ri B\omega}}{\omega-\ri} - \frac{1}{\omega+\ri} \biggr) \rd \omega = -\frac{m\re^B}{2} \frac{G_+(\ri)}{\sqrt{\pi}}\\
\quad{}\times\biggl[ -(2B)^{1/2}\frac{k}{G_+(\ri)}\frac{\sqrt{\pi}}{2}\biggl( 1 + \frac{\mathsf{K}}{\pi} \biggr)\frac{1}{x^{1/2}}
+ \frac{\sqrt{\pi}}{2} + \frac{\sqrt{\pi}}{G_+(\ri)}\sum_{n\ge 1} \frac{\ri\sigma^\pm_n G_+(\ri\xi_n)}{(\xi_n-1)\xi_n} \re^{-2B\xi_n} \biggr]\! + \mathcal{O}\bigl(B^0\bigr).\nonumber
\end{gather}

We can now write~\eqref{eq_QB_integral_equation} order by order in~the exponential corrections $\re^{-2B\xi_n}$, which yields the following integral equations for the functions $Q_{n}(x)$:
\begin{equation}
\biggl( 1 + \frac{\mathsf{K}}{\pi} \biggr) Q_{n}(x) = -\frac{m\re^B}{2} \frac{\ri\sigma^\pm_n G_+(\ri\xi_n)}{(\xi_n-1)\xi_n} - \frac{\ri\sigma^\pm_n Q(\ri\xi_n)}{\xi_n} + \mathcal{O}\bigl(B^0\bigr), \qquad n\ge 1.
\end{equation}
The solutions $Q_{n}(x)$ can be determined with the tools presented in~\cite[Appendix B]{mmr-an}. The case~${n=0}$ corresponds to the perturbative part, which is more involved, and is done in detail in~\cite[Appendix~A]{mmr-an}. In particular, the function $Q_{0}(x)$ has to be split in~terms proportional to the parameters $a$, $b$ appearing in~\eqref{eq_G_series}:
\begin{gather}
Q_{0}(x) = -kh \bigl[ B q_0(x) + a\log(2B)q_{2,1}^a(x) + a q_{2,0}^a(x) + b q_{2,0}^b(x) \bigr]\nonumber\\
\hphantom{Q_{0}(x) = }{}
 - \frac{m\re^B G_+(\ri)}{\sqrt{\pi}} \bigl[ (2B)^{1/2} q_1(x) + q_2(x) \bigr] + \mathcal{O}\bigl(B^0\bigr).
\end{gather}
We then find, for the integral involving $Q(\ri\xi')$ in~\eqref{eq_sigma_part},
\begin{gather}
\frac{1}{2\pi\ri} \int_0^\infty \frac{\re^{-2B\xi'}\delta\sigma(\ri\xi')Q(\ri\xi')}{\xi' + \xi_m} \rd\xi' = \frac{kh(2B)^{1/2}}{2\pi\xi_m}\nonumber\\[1mm]
\qquad{}\times\biggl[ \langle q_0 \rangle + \frac{1}{B}\biggl( a\log(2B) \bigl\langle q_{2,1}^a - xq_0\bigr\rangle+ a\bigl\langle q_{2,0}^a+x\log(x)q_0 \bigr\rangle + b\bigl\langle q_{2,0}^b + xq_0 \bigr\rangle - \frac{\langle xq_0\rangle}{2\xi_m} \biggr) \biggr]\nonumber\\[1mm]
\quad{}+ \frac{m\re^B}{\pi\xi_m} \frac{G_+(\ri)}{\sqrt{\pi}} \biggl[ \frac{\langle q_1 \rangle}{(2B)^{1/2}} + \frac{\langle q_2 \rangle}{2B} + \frac{\langle q_3 \rangle}{G_+(\ri)(2B)}\sum_{n\ge 1} \frac{\ri\sigma^\pm_n G_+(\ri\xi_n)}{(\xi_n-1)\xi_n}\re^{-2B\xi_n} \biggr]\nonumber\\[1mm]
\quad{}+ \frac{\langle q_3 \rangle}{\pi^{3/2}B\xi_m}\sum_{n\ge 1} \frac{\ri\sigma^\pm_n Q(\ri\xi_n)}{\xi_n} \re^{-2B\xi_n} + \mathcal{O}\bigl(B^{-1}\bigr),
\label{eq_Q_integral_part}
\end{gather}
where we have defined the momenta
\begin{equation}
\langle q \rangle = \int_0^\infty \re^{-x} q(x) \rd x.
\end{equation}
The perturbative part involves different momenta that are worked out in~\cite[Appendix B]{mmr-an}. We~have introduced the momentum of a new function $q_3(x)$, defined as
\begin{equation}
Q_{n}(x) = -\frac{2}{\sqrt{\pi}} \biggl( \frac{m\re^B}{2} \frac{\ri\sigma^\pm_n G_+(\ri\xi_n)}{(\xi_n-1)\xi_n} + \frac{\ri\sigma^\pm_n Q(\ri\xi_n)}{\xi_n} \biggr) q_3(x) + \mathcal{O}\bigl(B^0\bigr).
\end{equation}
Its momentum can be easily computed with the same tools as the perturbative part, and it is given by
\[
\langle q_3 \rangle = \pi^{3/2}/8.
\]

We now have to put all pieces together \eqref{eq_h_part_2},~\eqref{eq_m_part_2},~\eqref{eq_sigma_part},~\eqref{eq_Q_integral_part} in~\eqref{eq_Qm_integral_equation}. After evaluating the momenta and the integrals $I_0$, $I_1$, we obtain
\begin{gather}
Q(\ri\xi_m) = \frac{kh(2B)^{1/2}\sqrt{\pi}}{\xi_m} \biggl[ -\frac{1}{2} + \frac{-4 a (\log(2B)+\gamma -1+\log (4))+4 b-3/\xi_m}{16 B} \biggr]\nonumber\\[1mm]
\hphantom{Q(\ri\xi_m) =}{}
+s h\biggl[ \frac{G_+(\ri\xi_m)}{\xi_m} + \sum_{n\ge 1} \frac{\ri \sigma^\pm_n G_+(\ri\xi_n)}{(\xi_n+\xi_m)\xi_n} \re^{-2B\xi_n} \biggr]
\nonumber\\[1mm]
\hphantom{Q(\ri\xi_m) =}{}
-\frac{m\re^B}{2}\biggl[ \frac{G_+(\ri\xi_m)-G_+(\ri)}{\xi_m-1}+ \sum_{n\ge 1} \frac{\ri \sigma^\pm_n G_+(\ri\xi_n)}{(\xi_n+\xi_m)(\xi_n-1)} \re^{-2B\xi_n}\nonumber\\[1mm]
\hphantom{Q(\ri\xi_m) =-\frac{m\re^B}{2}\biggl[}{}
- \frac{1}{8B\xi_m}\biggl( G_+(\ri) + \sum_{n\ge 1}\frac{\ri\sigma^\pm_n G_+(\ri\xi_n)}{(\xi_n-1)\xi_n} \re^{-2B\xi_n} \biggr) \biggr]\nonumber\\[1mm]
\hphantom{Q(\ri\xi_m) =}{}
- \sum_{n\ge 1} \frac{\ri\sigma^\pm_n Q(\ri\xi_n)}{\xi_n+\xi_m} \re^{-2B\xi_n} + \frac{1}{8 B \xi_m} \sum_{n\ge 1} \frac{\ri \sigma^\pm_n Q(\ri\xi_n)}{\xi_n} \re^{-2B\xi_n} + \mathcal{O}\bigl( B^{-1} \bigr).
\label{eq_Qm_implicit}
\end{gather}
We introduced the parameter $s=1$ for convenience, in order to keep track of some of the terms in~the above expression. In both the boundary condition and the final result for the free energy, all contributions of terms proportional to $s$ cancel each other, thus we can set $s=0$ to simplify the computation.

On the other hand, we also need to compute $\epsilon_+(\ri\kappa)/G_+(\ri\kappa)$ from~\eqref{eq_epsilon_0}. The steps are mostly the same as we did for $Q(\ri\xi_m)$, but replacing $\xi_m$ by $-\kappa$. In~\eqref{eq_h_part_2} and~\eqref{eq_m_part_2}, contributions that arose from the pole at $-\ri\xi_m$ will be different, since the pole will now be in~the positive imaginary side, at $\ri\kappa$. The final result is then given by
\begin{gather}
\frac{\epsilon_+(\ri\kappa)}{G_+(\ri\kappa)} = -\frac{kh(2B)^{1/2}\sqrt{\pi}}{\kappa} \biggl[ -\frac{1}{2} + \frac{-4 a (\log(2B)+\gamma -1+\log (4))+4 b+3/\kappa}{16 B} \biggr]\nonumber\\
\hphantom{\frac{\epsilon_+(\ri\kappa)}{G_+(\ri\kappa)} =}{}
+s h\biggl[ - \delta_{\kappa,1} G_-(\ri)\re^{-2B} + \sum_{n\ge 1} \frac{\ri \sigma^\pm_n G_+(\ri\xi_n)}{(\xi_n-\kappa)\xi_n} \re^{-2B\xi_n} \biggr]
\nonumber\\
\hphantom{\frac{\epsilon_+(\ri\kappa)}{G_+(\ri\kappa)} =}{}
-\frac{m\re^B}{2}\biggl[\frac{G_+(\ri)}{\kappa+1} - \delta_{\kappa,1} G_-'(\ri)\re^{-2B}+ \sum_{n\ge 1} \frac{\ri \sigma^\pm_n G_+(\ri\xi_n)}{(\xi_n-\kappa)(\xi_n-1)} \re^{-2B\xi_n}\nonumber\\
\hphantom{\frac{\epsilon_+(\ri\kappa)}{G_+(\ri\kappa)} =-\frac{m\re^B}{2}\biggl[}{}
+ \frac{1}{8B\kappa}\biggl( G_+(\ri) + \sum_{n\ge 1}\frac{\ri\sigma^\pm_n G_+(\ri\xi_n)}{(\xi_n-1)\xi_n} \re^{-2B\xi_n} \biggr) \biggr]\nonumber\\
\hphantom{\frac{\epsilon_+(\ri\kappa)}{G_+(\ri\kappa)} =}{}
- \sum_{n\ge 1} \frac{\ri\sigma^\pm_n Q(\ri\xi_n)}{\xi_n-\kappa} \re^{-2B\xi_n} - \frac{1}{8 B \kappa} \sum_{n\ge 1} \frac{\ri \sigma^\pm_n Q(\ri\xi_n)}{\xi_n}\re^{-2B\xi_n} + \mathcal{O}\bigl( B^{-1} \bigr).
\label{eq_epsilon_kappa}
\end{gather}
For convenience, we only wrote the terms arising from the pole at $\ri\kappa$ for the particular cases $\kappa=1, +\infty$. In the latter case, these contributions vanish, as indicated by the Kronecker delta.

We are now in position to compute the relation between $m$ and $h$, resulting from the boundary condition~\eqref{eq_boundary_condition}. To this end, we propose the ansatz
\begin{align}
&m\re^B= kh (2B)^{1/2} \frac{\sqrt{\pi}}{G_+(\ri)} \sum_{n\ge 0} b_n (B) \re^{-2B\xi_n},\nonumber\\
&b_{n}(B)= b_{(n),0} + s\frac{b_{(n),1}}{\sqrt{B}} + \frac{b_{(n),2,1} \log(2B)}{B} + \frac{b_{(n),2,0}}{B} + \mathcal{O}\bigl(B^{-2}\bigr).
\label{eq_boundary_condition_ansatz}
\end{align}
We will see later that the $1/\sqrt{B}$ correction will be 0 to all exponential orders, so one can effectively set $s=0$ in~the expression for $b_n(B)$. To find the parameters in~the ansatz of $b_n(B)$, we first have to plug~\eqref{eq_boundary_condition_ansatz} in~\eqref{eq_Qm_implicit} and solve the resulting equation for the quantities $Q(\ri\xi_m)$, order by order in~the exponential corrections. We finally evaluate~\eqref{eq_epsilon_kappa} at $\kappa\rightarrow +\infty$, using our result for $Q(\ri\xi_m)$, and impose the boundary condition~\eqref{eq_boundary_condition} to solve for the parameters $b_{(n),0}$, $b_{(n),2,1}$, $b_{(n),2,0}$. As an example, for $n=1$, we obtain
\begin{gather}
b_1(B) = \frac{\ri \sigma^\pm_1}{(\xi_1 -1)\xi_1} + \frac{\ri \sigma^\pm_1 a \log(2B)}{2(\xi_1 - 1)\xi_1 B}
+\frac{\ri \sigma^\pm_1 (4 a (\gamma_E -1+\log (4))\xi_1 -4 b \xi_1-5 \xi_1+2)}{8 (\xi_1-1) \xi_1^2 B}.\!
\end{gather}

The free energy can be obtained by evaluating~\eqref{eq_epsilon_kappa} at $\kappa=1$ and putting the result in~\eqref{eq_free_energy_fourier}. In addition, we write instances of $m$ in~terms of $h$, by using~\eqref{eq_boundary_condition_ansatz} with the parameters obtained when we imposed the boundary condition. The final result for the free energy can be written as%
\begin{equation}
\mathcal{F}(h) = - \frac{m^2}{4\pi}\ri G_-'(\ri)G_+(\ri) - \frac{k^2 h^2 B}{4} \sum_{n\ge 0} f_n(B) \re^{-2B\xi_n},
\label{eq_free_energy_transseries_B}
\end{equation}
where the coefficients $f_n(B)$ are series in $1/B$ and $\log(2B)$. In particular, for the first exponential correction, we find
\begin{gather}
f_1(B) = -\frac{2\ri \sigma^\pm_1}{(\xi_1-1)^2 \xi_1} - \frac{2\ri\sigma^\pm_1 a \log(2B)}{(\xi_1-1)^2 \xi_1 B}\nonumber\\
\hphantom{f_1(B) =}{}
-\frac{\ri \sigma^\pm_1 \biggl(4 a (\gamma_E -1+\log (4))\xi_1 -4 b \xi_1+2 \xi_1^2-4 \xi_1+1\biggr)}{2 (\xi_1-1)^2 \xi_1^2 B}.
\end{gather}

In the next step, we will write the trans-series in~\eqref{eq_free_energy_transseries_B} as a trans-series in~the $\tilde\alpha$ coupling of~\eqref{eq_alpha_tilde_def}. A relation between $\tilde{\alpha}$ and the parameter $B$ can be extracted from the boundary condition derived in~\eqref{eq_boundary_condition_ansatz}. We propose the following trans-series ansatz
\begin{align}
&B= \frac{1}{\tilde{\alpha}} + \sum_{n\ge 0} d_n(\tilde{\alpha}) \tilde{\alpha}^{-2a\xi_n} \re^{-2\xi_n/\tilde{\alpha}},\nonumber\\
&d_n(\tilde{\alpha})= d_{(n),0} + d_{(n),2,1} \tilde{\alpha} \log\tilde{\alpha} + d_{(n),1}\tilde{\alpha} + \mathcal{O}\bigl(\tilde{\alpha}^2\bigr).
\label{eq_B_to_alpha_tilde}
\end{align}
The coefficients can then be obtained by combining~\eqref{eq_alpha_tilde_def} with~\eqref{eq_boundary_condition_ansatz}:
\begin{gather}
\frac{1}{\tilde{\alpha}} +\! \biggl( a + \frac{1}{2} \biggr)\log\tilde{\alpha} = B - \frac{1}{2}\log(B) + \frac{1}{2}\log\biggl( \frac{G_+(\ri)^2}{2\pi k^2} \frac{m^2}{\Lambda^2} \biggr)\! - \log\biggl( \sum_{n\ge 0} b_n(B) \re^{-2B\xi_n} \biggr).\!
\label{eq_alpha_def_mod}
\end{gather}
Replacing $B$ with the expression~\eqref{eq_B_to_alpha_tilde}, one obtains an equation for $d_k(\tilde{\alpha})$ that can be recursively solved for
\begin{gather}
d_k(\tilde{\alpha}) = \coeff_k \Biggl\{ \frac{1}{2} \log\Biggl[ 1 + \tilde{\alpha} \sum_{i=0}^{k-1} d_i(\tilde{\alpha}) \tilde{\alpha}^{-2a\xi_i} \re^{-2\xi_i/\tilde{\alpha}}\Biggr] \\
\hphantom{d_k(\tilde{\alpha}) = \coeff_k \Biggl\{}{}
+ \log\Biggl[ \sum_{i=0}^{k-1} \biggl( \frac{G_+(\ri)^2}{2\pi k^2} \frac{m^2}{\Lambda^2} \biggr)^{\xi_i} b_i(\tilde{\alpha})\exp\Biggl( -2\xi_i \sum_{j=0}^{k-1} b_j(\tilde{\alpha}) \tilde{\alpha}^{-2a\xi_j} \re^{-2\xi_j/\tilde{\alpha}} \Biggr) \Biggr] \Biggr\},\nonumber
\end{gather}
where $\coeff_k(f)$ denotes the $k$-th coefficient of $f$ as an expansion in powers of $y=\tilde{\alpha}^{-2a\xi_1} \re^{-2\xi_1/\tilde{\alpha}}$, and $b_j(\tilde{\alpha})$ are the coefficients $b_j(B)$ evaluated at $B=1/\tilde{\alpha}$.\footnote{This replacement is only valid at the order we are working.} The initial condition can also be extracted from~\eqref{eq_alpha_def_mod}, with the result
\begin{gather}
d_0(\tilde{\alpha}) = a \log \tilde{\alpha} - \frac{1}{2}\log\biggl( \frac{G_+(\ri)^2}{2\pi k^2} \frac{m^2}{\Lambda^2} \biggr)\nonumber\\
\hphantom{d_0(\tilde{\alpha}) =}{}
+ \frac{4 a (\gamma_E -1+\log (8))-4 b-2 \log\bigl( \frac{G_+(\ri)^2}{2\pi k^2} \frac{m^2}{\Lambda^2} \bigr)-1}{8} \tilde{\alpha} + \mathcal{O}\bigl(\tilde{\alpha}^2\bigr).
\end{gather}

We can now write the free energy $\mathcal{F}(h)$ that we determined in~\eqref{eq_free_energy_transseries_B} as a trans-series in $\tilde{\alpha}$, which yields the result of~\eqref{eq_free_energy_transseries}. All contributions of $\log \tilde\alpha$ cancel in~this step and the $\varphi_\ell(\tilde{\alpha})$ are simply power series in $\tilde{\alpha}$.

The coefficient $\mathcal{C}_\text{IR}^\pm$ in~\eqref{eq_free_energy_transseries} can be extracted by rewriting the $m^2$ term of~\eqref{eq_free_energy_transseries_B} in~terms of~$h^2$, directly using the coupling definition~\eqref{eq_alpha_tilde_def}. One obtains
\begin{equation}
\mathcal{C}_\text{IR}^\pm = \frac{\ri G_-'(\ri) G_+(\ri)}{\pi k^2} \frac{m^2}{\Lambda^2}.
\end{equation}
The ambiguity in~this coefficient arises from the factor $G_-'(\ri)$.

\subsubsection{Numerical analysis}
Using the method discussed in~the previous section, we can now generate the first two coefficients of the series
\begin{equation}
\varphi_\ell(\tilde\alpha) = a_{0,\ell} + a_{1,\ell} \tilde\alpha + \mathcal{O}\bigl(\tilde\alpha^2\bigr)
\end{equation}
appearing in~the trans-series~\eqref{eq_free_energy_transseries}, up to very large order in $\ell$. To the very first values of $\ell$, we find the following analytic results, valid for generic bosonic models:
\begin{gather}
\varphi_1(\tilde{\alpha}) = \biggl( \frac{G_+(\ri)^2}{2\pi k^2} \frac{m^2}{\Lambda^2} \biggr)^{\xi_1} \biggl[ -\frac{2 \ri \sigma_1^\pm}{(\xi_1-1)^2 \xi_1}\label{eq_coeff_bosonic_1}\\
\hphantom{\varphi_1(\tilde{\alpha}) = }{}
+\frac{\ri \sigma_1^\pm \bigl(4 a (\gamma_E -1+\log (8))\xi_1 -2\xi_1 \bigl( 4 b +\log\bigl( \frac{G_+(\ri)^2}{2\pi k^2} \frac{m^2}{\Lambda^2} \bigr)\bigr) -\xi_1 +1\bigr)}{2 (\xi_1-1) \xi_1^2} \tilde{\alpha} + \mathcal{O}\bigl(\tilde{\alpha}^2\bigr) \biggr],\nonumber
\\
 \varphi_2(\tilde\alpha) =\biggl( \frac{G_+(\ri)^2}{2\pi k^2} \frac{m^2}{\Lambda^2} \biggr)^{2\xi_1} \biggl[ \frac{2 \bigl(\ri\sigma_1^\pm\bigr)^2}{(\xi_1-1)^2 \xi_1^2}-\frac{\ri \sigma_2^\pm}{(2 \xi_1-1)^2 \xi_1}\nonumber\\
 + \biggl(\!- \frac{\hspace{-0.7pt}\bigl(\ri\sigma_1^\pm\bigr)^2 \hspace{-0.7pt}\bigl(2 \xi_1 (2 a (2 \xi_1\hspace{-1pt}-\hspace{-0.7pt}1) (\gamma_E \hspace{-1pt}-\hspace{-0.7pt}1\hspace{-0.7pt}+\hspace{-0.7pt}\log (8))\hspace{-0.7pt} -\hspace{-0.7pt} (2\xi_1\hspace{-1pt}-\hspace{-0.7pt}1)\bigl(4 b\hspace{-0.7pt}+\hspace{-0.7pt}\log\bigl( \frac{G_+(\ri)^2}{2\pi k^2} \frac{m^2}{\Lambda^2} \bigr) \bigr) \hspace{-1pt}-\hspace{-0.7pt} \xi_1 \hspace{-1pt}+ \hspace{-0.7pt}1)\hspace{-0.7pt}-\hspace{-0.7pt}1\bigr)}{2(\xi_1\hspace{-1pt}-\hspace{-0.7pt}1)^2\xi_1^3}\nonumber\\
+ \frac{\ri \sigma_2^\pm \bigl(2 \xi_1 \bigl(4 a (\gamma_E -1+\log (8))-8 b-2 \log\bigl( \frac{G_+(\ri)^2}{2\pi k^2} \frac{m^2}{\Lambda^2} \bigr)-1\bigr)+1\bigr)}{8 (2\xi_1-1)\xi_1^2}\biggr) \tilde\alpha + \mathcal{O}\bigl(\tilde\alpha ^2\bigr) \biggl], \label{eq_coeff_bosonic_2}
\\
\varphi_3(\tilde\alpha) = \biggl( \frac{G_+(\ri)^2}{2\pi k^2} \frac{m^2}{\Lambda^2} \biggr)^{3\xi_1} \biggl[ -\frac{(7 \xi_1-3) \bigl(\ri\sigma_1^\pm\bigr)^3}{2(\xi_1-1)^3\xi_1^3} + \frac{2 \bigl(\ri\sigma_1^\pm\bigr) \bigl(\ri\sigma_1^\pm\bigr)}{(\xi_1-1)(2\xi_1-1)\xi_1^2} - \frac{2 \ri \sigma_3^\pm}{3(3 \xi_1-1)^2\xi_1}\!\!\label{eq_coeff_bosonic_3}\\
\qquad{}
+ \biggl( -\frac{ \bigl(\ri\sigma_1^\pm\bigr)^3 \bigl(4b + \log\bigl( \frac{G_+(\ri)^2}{2\pi k^2} \frac{m^2}{\Lambda^2} \bigr) \bigr) (3 \xi_1-1) (7 \xi_1-3)}{4(\xi_1-1)^3 \xi_1^3}\nonumber\\
\qquad\quad{}
+\frac{ \bigl(\ri\sigma_1^\pm\bigr)^3 \bigl(\xi_1 \bigl(12 a (3 \xi_1-1) (7 \xi_1-3) (\gamma_E -1+\log (8))-63 \xi_1^2+69 \xi_1-49\bigr)+3\bigr)}{24 (\xi_1-1)^3 \xi_1^4}\nonumber\\
\qquad\quad{}
-\frac{\bigl(\ri\sigma_1^\pm\bigr) \bigl(\ri\sigma_2^\pm\bigr) \bigl(4 b + \log\bigl( \frac{G_+(\ri)^2}{2\pi k^2} \frac{m^2}{\Lambda^2} \bigr) \bigr) (1-3 \xi_1)}{(\xi_1-1)(2 \xi_1-1) \xi_1^2 }\nonumber\\
\qquad\quad{}
- \frac{\bigl(\ri\sigma_1^\pm\bigr) \bigl(\ri\sigma_2^\pm\bigr) \bigl(18 \xi_1^2 (4 a (\gamma_E -1+\log (8))-1)+12 \xi_1 (1-2 a (\gamma_E -1+\log (8)))-5\bigr)}{12 (\xi_1-1) (2 \xi_1-1) \xi_1^3}\nonumber\\
\qquad\quad{}
+ \frac{\ri \sigma_3^\pm \bigl(3 \xi_1 \bigl(4 a (\gamma_E -1+\log (8))-8 b - 2 \log\bigl( \frac{G_+(\ri)^2}{2\pi k^2} \frac{m^2}{\Lambda^2} \bigr) -1\bigr)+1\bigr)}{18 \xi_1^2 (3 \xi_1-1)} \biggr) \alpha + \mathcal{O}\bigl(\alpha^2\bigr)
\biggr].\nonumber
\end{gather}

To extract the large order behavior of the coefficients $a_{n,\ell}$, at large $\ell$, we will do a numerical computation, focusing on the ${\rm O}(N)$ non-linear sigma model and the ${\rm SU}(N)$ PCF. In the following, we summarize all the relevant parameters that we need for each of the two models.
\begin{enumerate}\itemsep=0pt
\item[(i)] \textit{Non-linear ${\rm O}(N)$ sigma model.} With the choice of charges $\mathsf{Q}$ made in~\cite{hmn,hn}, the function~$G_+(\omega)$ that one obtains from the Wiener--Hopf decomposition of the kernel is given~by%
\begin{equation}
G_+(\omega) = \frac{\re^{-{1 \over 2} \ri \omega [(1-2\Delta)(\log(-\frac{1}{2}\ri \omega)-1)-2\Delta\log(2\Delta)]}}{\sqrt{-\ri \Delta \omega}}\frac{\Gamma(1-\ri \Delta \omega)}{\Gamma\bigl(\tfrac{1}{2}-\tfrac{1}{2}\ri\omega\bigr)}, \qquad \Delta = \frac{1}{N-2}.
\end{equation}
The parameters of the expansion~\eqref{eq_G_series} can be easily extracted from the above expression, obtaining
\begin{gather}
k = \frac{1}{\sqrt{\pi\Delta}}, \qquad a = -\frac{1}{2} (1-2 \Delta), \nonumber\\ b=\frac{(\gamma_E -1) (2 \Delta -1)+2 \Delta \log (\Delta )-\log (2)}{2}.
\end{gather}
The singularities and residues of $\sigma(\omega)$, defined in~\eqref{eq_sigma_def}, are\footnote{When $N$ is odd, there are singularities only for even $\ell$ in~\eqref{eq_ON_singularities}. We can still treat all values of $N$ with the same formulas, with the observation that $\ri\sigma_\ell^\pm = 0$ when there is no actual singularity.}
\begin{equation}
\xi_\ell = \frac{\ell}{\Delta}, \qquad \ri\sigma_\ell^\pm = \pm \ri \frac{ \bigl( \frac{\ell}{\re}\bigr)^{2 \ell} \bigl(\frac{\ell}{2\ri\re\Delta }\bigr)^{-\frac{\ell}{\Delta }} }{\Delta (\ell-1)!\, \ell! } \frac{\Gamma \bigl(\frac{1}{2} + \frac{\ell}{2 \Delta }\bigr)}{\Gamma \bigl(\frac{1}{2} - \frac{\ell}{2 \Delta }\bigr)}.
\label{eq_ON_singularities}
\end{equation}
The relation between the mass gap and the dynamically generated scale in~the $\overline{\text{MS}}$ scheme is given by~\cite{hmn,hn}
\begin{equation}
\frac{m}{\Lambda} = \biggl(\frac{8}{\re}\biggr)^{\Delta } \frac{1}{\Gamma (\Delta +1)}.
\end{equation}

\item[(ii)] \textit{Principal chiral field}. With the charges chosen in~\cite{pcf}, we have
\begin{gather}
G_+(\omega)=\frac{\re^{- \ri \omega [-(1-\Delta) \log (1-\Delta)-\Delta \log (\Delta )]}}{\sqrt{-2 \pi\ri (1-\Delta ) \Delta \omega } }\frac{\Gamma (1-\ri (1-\Delta ) \omega ) \Gamma (1-\ri \Delta \omega ) }{\Gamma (1-\ri \omega )},
\end{gather}
where $\Delta = \frac{1}{N}$.
The parameters of the expansion~\eqref{eq_G_series} are
\begin{equation}
k = \frac{1}{\sqrt{2\pi (1-\Delta)\Delta}}, \qquad a = 0, \qquad b = (1-\Delta ) \log (1-\Delta )+\Delta \log (\Delta ).
\end{equation}
The singularities and residues of $\sigma(\omega)$ are
\begin{gather}
\xi_\ell = \frac{\ell}{1-\Delta}, \nonumber\\
 \ri\sigma_\ell^\pm = \pm\ri \frac{(-1)^{\ell+1} (1-\Delta )^{2 \ell} \Delta ^{\frac{2 \Delta \ell}{1-\Delta}} }{\ell!^2} \frac{\Gamma \bigl(1+\frac{\ell}{1-\Delta }\bigr) \Gamma \bigl(1+\frac{\ell \Delta }{\Delta -1}\bigr)}{\Gamma \bigl(-\frac{\ell}{1-\Delta}\bigr) \Gamma \bigl(1+\frac{\ell \Delta }{1-\Delta }\bigr)}.
\end{gather}
The mass gap is given by~\cite{pcf}
\begin{equation}
\frac{m}{\Lambda} = \sqrt{\frac{8 \pi }{\re}} \frac{\sin (\pi \Delta )}{\pi \Delta }.
\end{equation}
\end{enumerate}

We will check, up to numerical error, and for each of the two models discussed above, that the coefficients $a_{n,\ell}$ appearing in~the trans-series have the asymptotics~\eqref{an-bn-R}. To this end, we~construct the quantity
\begin{equation}
\biggl| \frac{a_{n,\ell}}{a_{n,\ell+1}} \biggr| = R \biggl( 1 + \frac{b_n}{\ell} + \mathcal{O}\bigl(1/\ell^2\bigr) \biggr).
\label{eq_anl_quotient}
\end{equation}
The coefficients $a_{n,\ell}$ are ambiguous, due to the two choices in~the residues \smash{$\ri\sigma_\ell^\pm$}. These two choices differ by complex conjugation: \smash{$\bigl( \ri\sigma_\ell^+ \bigr)^ *=\ri \sigma_\ell^-$}. On the other hand, as seen explicitly in~the expressions of~\eqref{eq_coeff_bosonic_1}--\eqref{eq_coeff_bosonic_3}, the $a_{n,\ell}$s are multinomials of $\ri\sigma_\ell^\pm$, with real coefficients. Therefore,
the two choices gives the same coefficients $a_{n, \ell}$, up to complex conjugation. This implies that $|a_{n,\ell}|$ is not ambiguous and, thus, neither are the parameters $R$ and $b_n$ appearing in~\eqref{eq_anl_quotient}.\looseness=-1

\begin{figure}[t]
\centering
\includegraphics[width=0.6\textwidth]{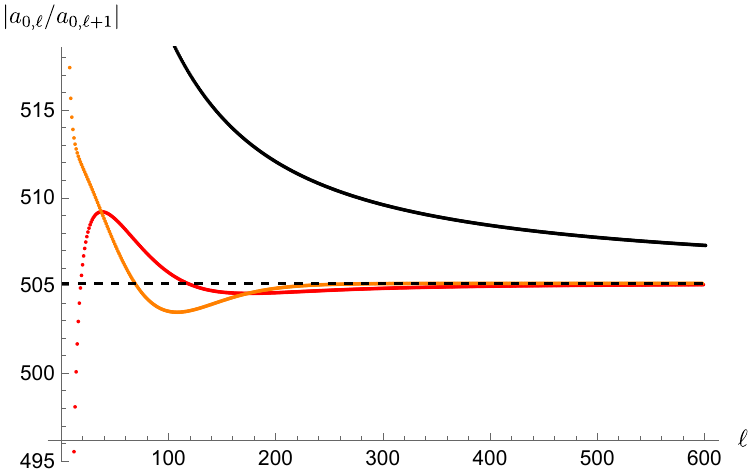}
\caption{Sequence $| a_{0,\ell}/a_{0,\ell+1} |$ for the free energy of the ${\rm O}(8)$ non-linear sigma model (black), together with the 1st (red) and 2nd (orange) Richardson transforms. The dashed line is the value $R$ to which the sequences converge.}
\label{fig_radius_ON}
\end{figure}

Using Richardson transforms, one can accelerate the convergence of the sequence when $\ell$ is large, extracting the radius of convergence $R$. In Figure~\ref{fig_radius_ON}, we show the result for the ${\rm O}(8)$ non-linear sigma model and $n=0$.

\begin{figure}[t]
\centering
\includegraphics[width=0.6\textwidth]{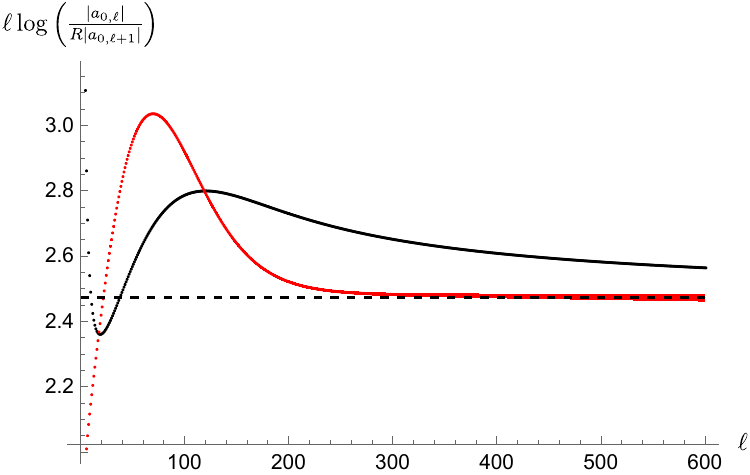}
\caption{Sequence \smash{$\ell \log\bigl( \frac{| a_{0,\ell} |}{R | a_{0,\ell+1} |}\bigr)$} for the free energy of the ${\rm O}(8)$ non-linear sigma model (black), together with the 1st Richardson transform (red). The dashed line is the value $b_0$ to which the sequences converge.}
\label{fig_b_ON}
\end{figure}

Further, we can extract the value of $b_n$ by considering the sequence{\samepage
\begin{equation}
\label{bn-seq}
\ell \log\biggl( \frac{| a_{n,\ell} |}{R | a_{n,\ell+1} |}\biggr) = b_n + \mathcal{O}(1/\ell).
\end{equation}
In Figure \ref{fig_b_ON}, we show the result for the ${\rm O}(8)$ non-linear sigma model and $n=0$.}

In Table~\ref{tab_ON}, we display the value of the parameters $R$ and $b_n$ that we
obtained for the ${\rm O}(N)$ non-linear sigma model, at different values of $N$ and for the leading
$n=0$ and subleading~${n=1}$ coefficients in~the trans-series. In Table~\ref{tab_PCF}, we show the same
results for the ${\rm SU}(N)$ PCF model. We find numerical evidence that the radius of convergence
$R$ does not depend on $n$. On~the other hand, the numeric results seem to indicate that $b_1 = b_0 - 1$. We also note
that, for the ${\rm O}(N)$ sigma model, when $N$ is even, the result for $b_0$ is compatible with the exact result~${b_0=5/2}$, which is also the value found in~the
Gross--Neveu model. Our numerical analysis uses 600 coefficients $a_{n,\ell}$, for both $n=0$ and $n=1$. The
numerical errors of $R$ are estimated from the difference between two successive
Richardson transforms, and we perform a number of Richardson transforms so that the
error estimate is minimized. In the case of $b_n$, the error estimate is the error propagating from
$R$ plus the error estimated from two successive Richardson transforms, with both errors added in quadrature.

\begin{table}[t]\centering
\caption{Parameters $R$ and $b_n$ of the asymptotics~\eqref{an-bn-R}, for the free energy of the ${\rm O}(N)$ non-linear sigma model at different values of $N$. We extract the parameters from the leading $n=0$ and subleading $n=1$ corrections in $\tilde\alpha$. For the radius $R$, we find numerical evidence that this parameter does not depend on~$n$.}\label{tab_ON}\vspace{-4mm}
$$
\arraycolsep=3mm
\begin{array}{lllll}
\toprule
N & R \text{ leading} & R \text{ subleading} & b_0 & b_1 \\
\midrule
4 & 0.25294249(91\pm 15) & 0.25294250(04\pm 26) & 2.5(00\pm 12) & 1.5(00\pm 15) \\
5 & 2.10247819(7\pm 7) & 2.10247819(7\pm 6) & 2.8(28\pm 12) & 1.8(28\pm 10) \\
6 & 9.30450(59\pm 13) & 9.30450(60\pm 13) & 2.(46\pm 11) & 1.(46\pm 11)\\
7 & 89.848306(72\pm 19) & 89.848306(72\pm 18) & 2.8(29\pm 11) & 1.82(9\pm 9) \\
8 & 505.12143(7\pm 9) & 505.12143(8\pm 8) & 2.4(7\pm 9) & 1.4(7\pm 9)\\
9 & 5427.95182(8\pm 5) & 5427.95182(8\pm 5) & 2.8(30\pm 10) & 1.83(0\pm 9) \\
10 & 34063.91(7\pm 4) & 34063.91(7\pm 4) & 2.(48\pm 11) & 1.(48\pm 11) \\
11 & 388517.904(71\pm 13) & 388517.904(71\pm 12) & 2.8(30\pm 10) & 1.83(0\pm 8) \\
\bottomrule
\end{array}
$$
\end{table}

\begin{table}[t]\centering
\caption{Parameters $R$ and $b_n$ of the asymptotics~\eqref{an-bn-R}, for the free energy of the ${\rm SU}(N)$ PCF model at different values of $N$.}\label{tab_PCF}\vspace{-4mm}
$$
\arraycolsep=3mm
\begin{array}{lllll}
\toprule
N & R \text{ leading} & R \text{ subleading} & b_0 & b_1 \\
\midrule
2 & 0.25294249(91\pm 15) & 0.25294250(04\pm 26) & 2.4(96\pm 18) & 1.50(0\pm 8) \\
3 & 0.17492(41\pm 29) & 0.17492(73\pm 10) & 2.5(80\pm 29) & 1.4(99\pm 32) \\
4 & 0.16537769(3\pm 6) & 0.16537769(4\pm 6) & 2.8(57\pm 30) & 1.8(57\pm 15)\\
5 & 0.1489842(2\pm 7) & 0.1489842(2\pm 8) & 2.8(80\pm 32) & 1.8(80\pm 33) \\
6 & 0.149513(8\pm 8) & 0.149513(6\pm 9) & 2.8(84\pm 35) & 1.8(83\pm 21)\\
7 & 0.15488(63\pm 30) & 0.15488(64\pm 31) & 2.8(4\pm 4) & 1.8(33\pm 22) \\
8 & 0.15(78\pm 23) & 0.15(81\pm 18) & 6\pm 11 & 4\pm 8 \\
9 & 0.15671(89\pm 14) & 0.15671(44\pm 11) & 2.9(1\pm 5) & 1.92(9\pm 6) \\
\bottomrule
\end{array}
$$
\end{table}

\subsection{The Gross--Neveu model}
\label{sec_gross_neveu}

The Gross--Neveu model is an integrable model of $N$ Majorana fermions in 1+1 dimensions with a quartic interaction.
We restrict ourselves to the case $N>4$. Its Lagrangian is given by
\begin{equation}
\CL = \frac{\ri}{2}\bar{\boldsymbol{\chi}}\cdot\slashed{\partial}\boldsymbol{\chi}+\frac{g^2}{8}\left(\bar{\boldsymbol{\chi}}\cdot\boldsymbol{\chi}\right)^2.
\end{equation}
This model is integrable and its $S$-matrix is known exactly. As in (\ref{eq_hamiltonian_mod}), we
will introduce a~chemical potential $h$ coupled to a conserved charge $\mathsf{Q}$, which will be
chosen as in~\cite{fnw1,fnw2}. In~the Gross--Neveu case, the Wiener--Hopf formalism can be
slightly simplified, and as described in~\cite{fnw1,mmr-an}, one can encode the solution to~\eqref{eq_Q_integral_equation} in a function $u(\omega)$
that solves the integral equation
\begin{equation}
u(\omega) = \frac{\ri}{\omega}+ \frac{1}{2\pi\ri}\int_\IR \frac{\re^{2\ri B \omega'}\rho(\omega')u(\omega') }{\omega'+\omega+\ri 0} \rd \omega'.
\label{inteq_u}
\end{equation}
The boundary condition~\eqref{eq_boundary_condition_0} reads in~this case
\begin{equation}
u(\ri) = \frac{m \re^B}{2h} \frac{G_+(\ri)}{G_+(0)}.
\label{u_bc}
\end{equation}
The
function $\rho(\omega)$ in~\eqref{inteq_u} is defined as
\begin{equation}
\rho(\omega) = - \frac{\omega+\ri}{\omega-\ri}\sigma(\omega),
\label{rhofunc_def}
\end{equation}
where $\sigma(\omega)$ was introduced in~\eqref{eq_sigma_def}. The Wiener--Hopf decomposition of the kernel gives
\begin{equation}
G_+(\omega)=\re^{-\frac{\Upsilon}{2}\ri\omega[1-\log(-\frac{\Upsilon}{2}\ri\omega)]+\frac{1}{2}\ri\omega[1-\log(-\frac{1}{2}\ri\omega)]} \frac{\Gamma\bigl(\frac{1}{2} - \frac{\Upsilon}{2}\ri\omega\bigr)}{\Gamma\bigl(\frac{1}{2} - \frac{1}{2}\ri\omega\bigr)},\qquad \Upsilon = \frac{N-4}{N-2}.
\label{GN_Gplus_usec}
\end{equation}
Lastly, the free energy (\ref{subF}) can be obtained directly from $u(\omega)$ through
\begin{equation}
\CF(h) = - \frac{h^2}{2\pi}u(\ri)G_+(0)^2\biggl(1- \frac{1}{2\pi\ri}\int_\IR \frac{\re^{2\ri B \omega'}\rho(\omega')u(\omega') }{\omega'-\ri} \rd \omega'\biggr).
\label{fh-u}
\end{equation}

%
Using the techniques described in~\cite{mmr-an}, which we outline in Section~\ref{gn-sum},
one can extract the trans-series expansion of the free energy from the integral equation.
As found therein, the trans-series can be written as
\begin{equation}
\Phi_\pm({\alpha}) = \varphi_0(\tilde{\alpha})+ \CC_{\text{IR}}^\pm \re^{-\frac{2}{\tilde{\alpha}}} \tilde{\alpha}^{2\Delta}
+ \sum_{\ell\geq 1} \CC_\ell^\pm \re^{-\frac{2}{1-2\Delta}\frac{\ell}{\tilde{\alpha}}}\tilde\alpha^{\frac{2\Delta}{1-2\Delta}\ell}\varphi_\ell(\tilde{\alpha}).
\label{free-energy-ts}
\end{equation}
The observable itself is then given by Borel summation,
\begin{equation}
\CF(h) = - \frac{h^2}{2\pi} s_\pm(\Phi_\pm)(\tilde{\alpha})= - \frac{h^2}{2\pi}\biggl\{\sum_{\ell\geq 0} y^\ell \CC_\ell^\pm s_\pm(\varphi_\ell)(\tilde{\alpha}) + \CC_{\text{IR}}^\pm \re^{-\frac{2}{\tilde{\alpha}}} \tilde{\alpha}^{2\Delta}\biggr\},\label{sum-kappa}
\end{equation}
where
\begin{equation}
\CC^\pm_0=1,\qquad
y = \re^{-\frac{2}{\Upsilon\tilde{\alpha}}}\tilde\alpha^{\frac{2\Delta}{\Upsilon}},\qquad \Upsilon = 1-2\Delta,\qquad \Delta = \frac{1}{N-2}.
\end{equation}
This trans-series is expressed in~terms of the auxiliary coupling defined as
\begin{equation}
\frac{1}{\tilde{\alpha}}-\Delta\log\tilde{\alpha}=\log\frac{h}{\Lambda},
\label{tilde-alpha-def}
\end{equation}
where $\Lambda$ is the dynamically generated scale. This scale can be related to the non-perturbative mass of the model through the mass gap derived in~\cite{fnw1,fnw2},
\begin{equation}
\frac{m}{\Lambda} = \frac{(2\re)^\Delta}{\Gamma(1-\Delta)}.
\label{eq_mass_gap_GN}
\end{equation}
As previously outlined, the key test is to reorganize the asymptotic series $\varphi_{\ell}(\tilde{\alpha})$
and consider the radius of convergence of the series $\CG_n(y)$ indexed by $\tilde{\alpha}^n$.
As in (\ref{eq_free_energy_transseries}), the single, distinct, IR renormalon term proportional to
\smash{$\re^{-\frac{2}{\tilde{\alpha}}} \tilde{\alpha}^{2\Delta}$} is not relevant for the analysis of the trans-series convergence.

\subsubsection{Summary of the recursive problem}
\label{gn-sum}

In contrast to the bosonic models studied in~the previous section, the trans-series expansion of the integral equation
of the Gross--Neveu model can be cast into a simple recursive problem. Here, we summarize its key
ingredients, originally presented in~\cite{mmr-an}.

We introduce the variable $\eta$ and the coupling $v$ such that
\begin{equation}
\frac{1}{v}-2\Delta\log v = 2B, \qquad \omega = \ri v\eta.
\label{v-def}
\end{equation}
For the double expansion, it is convenient to define the trans-monomial
\begin{equation}
q= \re^{-1/(\Upsilon v)}v^{2\Delta/\Upsilon}.
\end{equation}
We then organize the expansion of $u(\eta) = u(\omega=\ri v \eta)$ as
\begin{equation}
u(\eta) = \sum_{\ell\geq 0} \sum_{n\geq 0}
 \bigl(\re^{\mp\frac{\ri\pi\Delta}{\Upsilon}}\bigr)^{\ell} q^\ell v^{n-1} u^{(n,\ell)}(\eta).
\end{equation}
It is convenient to include the phase in~the non-perturbative coupling,
\begin{equation}
\bigl(\re^{\mp\frac{\ri\pi\Delta}{\Upsilon}}\bigr)^{\ell} q^\ell = q_\pm^\ell.
\end{equation}

The trans-series solution for $u$ can be found with a ``seed solution'' and an iterated integral operator. The seed solution is
\be
\bar{u}(\eta) = \frac{1}{v\eta} + \sum_{n\geq 1}
q^{2n-1}_\pm
\frac{\bar{\rho}_n u_n}{\xi_n+v\eta},
\label{seed-solution}
\ee
where
\begin{equation}
u_k = u(\omega = \ri \xi_k),\qquad \xi_k = \frac{2k-1}{\Upsilon},
\end{equation}
and the $\bar\rho_n$ are real numbers,
\begin{equation}
\bar\rho_n=\frac{2}{\Upsilon} \frac{(-1)^{n}}{\Gamma(n)^2} \biggl(\frac{2n-1}{2\re}\biggr)^{2n-1} \biggl( \frac{2n-1}{2\Upsilon\re} \biggr)^{-\frac{2n-1}{\Upsilon}} \frac{\Gamma\bigl(\frac{3}{2} + \frac{2n-1}{2\Upsilon}\bigr)}{\Gamma\bigl(\frac{3}{2} - \frac{2n-1}{2\Upsilon}\bigr)},
\label{polepm}
\end{equation}
derived from the residues of $\rho(\omega)$,
\begin{equation}
\text{Res}_{\omega=\ri\xi_n\pm 0}\, \rho(\omega) = \bigl(\re^{\mp\frac{\ri\pi\Delta}{\Upsilon}}\bigr)^{2n-1}\ri \bar\rho_n.
\label{rho_factor}
\end{equation}
The expansion in $v$ can be solved by iterating the integral kernel $\mathfrak{I}$,
\begin{equation}
[\mathfrak{I} f](\eta) = - \frac{1}{\pi}\int_0^\infty \frac{\re^{-\eta'}P(\eta')}{\eta+\eta'} f(\eta')\rd\eta',
\end{equation}
where the kernel $P(\eta)$ is defined from the discontinuity of $\rho$,
\begin{equation}
\rho\bigl(\re^{-\ri 0}\omega\bigr)-\rho\bigl(\re^{+\ri 0}\omega\bigr) = -2\ri v \re^{2B\xi} \re^{-\eta}P(\eta),
\label{P-def}
\end{equation}
and can be asymptotically expanded,
\begin{equation}
 \quad P(\eta) \approx \sum_{n\geq 0} v^n \sum_{m=0}^n d_{n+1,m} \eta^{n+1} (\log\eta)^m\sim d_{1,0}\eta + \CO(v).
 \label{P-expansion}
\end{equation}
The solution to the integral equation can then be organized as
\begin{equation}
u(\eta)= \sum_{m\geq 0} v^m [\mathfrak{I}^m\bar{u}](\eta) = (1-v\mathfrak{I})^{-1}\bar{u}(\eta) .
\end{equation}
And we can also write the auxiliary functions
\begin{align}
&\bar{u}(\eta)= \sum_{\ell\geq 0} q_\pm^\ell \bar{u}^{(\ell)}(\eta),\nonumber\\
&u^{(\ell)}(\eta)= (1-v\mathfrak{I})^{-1}\bar{u}^{(\ell)}(\eta) = \sum_{n\geq 0} v^{n-1} u^{(n,\ell)}(\eta).
\label{ueta}
\end{align}
Note that $u^{(n,\ell)}(\eta)$ is distinct from $\mathfrak{I}^n\bar{u}^{(\ell)}$ because both $\bar{u}(\eta)$ and the integral operator $\mathfrak{I}$ produce higher corrections in $v$.

The solution is not yet fully specified, because we still need to find the values $u_k$ in~\eqref{seed-solution}.
Defining the integrals
\begin{equation}
\mathfrak{I}_{k} f = [\mathfrak{I} f]\bigl(\tfrac{2k-1}{v\Upsilon}\bigr),\qquad k\in\IN,
\end{equation}
we can write $u_k$ as
\begin{equation}
u_k=\frac{\Upsilon}{2k-1}+v \mathfrak{I}_k u + \frac{\Upsilon}{2}\sum_{n\geq 1}q^{2n-1}_\pm\frac{\bar\rho_{n} u_n}{k+n-1},\qquad k\in\IN.
\label{eq_uk_v2}
\end{equation}
Following the notation of~\eqref{ueta}, we define the expansions
\begin{equation}
u_k = \sum_{\ell \geq 0} q_\pm^\ell u_k^{(\ell)},\qquad
u^{(\ell)}_k = \sum_{n\geq 0} v^{n} u^{(n,\ell)}_k.
\end{equation}
This leads to a solvable recursive problem, as presented in~\cite{mmr-an},
\begin{align}
&u_k^{(0)}=\frac{\Upsilon}{2k-1}+v\mathfrak{I}_k(1-v\mathfrak{I})^{-1}{\bar u^{(0)}}\label{uks-seed},\\
&u_k^{(s)}=v \mathfrak{I}_k(1-v\mathfrak{I})^{-1}{\bar u^{(s)}} + \frac{\Upsilon}{2}\sum_{\tfrac{s+1}{2}\geq r\geq 1} \frac{\bar\rho_r u_r^{(s+1-2r)}}{k+r-1}\label{uks-recurs}.
\end{align}
The right-hand side does not contain \smash{$u^{(s)}_k$}, only \smash{$u_{k'}^{(s')}$} with $s<s'$ and \smash{$k'\leq \frac{s+1}{2}$}. The \smash{$u_k^{(n,m)}$} can be obtained by expanding in $v$ the operator $\mathfrak{I}_k$ as well as $\bar{u}^{(s)}$ itself.

An important intermediate ingredient is $u(\omega=\ri)$, which we denote by $u_\ri$. It can be calculated, using the above ingredients, from the equation:
\begin{equation}
u_\ri=1+v \bigl[\mathfrak{I} (1-v\mathfrak{I})^{-1} \bar{u}\bigr]\bigl(v^{-1}\bigr) + \Upsilon\sum_{n\geq 1}
q^{2n-1}_\pm
\frac{\bar\rho_{n} u_n}{\Upsilon+2n-1}.
\label{eq_ui_v2}
\end{equation}
We will need this value to compute $\CF(h)$, and it is also central to the change of variables from~$v$ to~$\tilde{\alpha}$.

To compute the free energy let us define the ``bare energy'',
\begin{equation}
f_v = 1 - v \bigl[\mathfrak{I} (1-v\mathfrak{I})^{-1} \bar{u}\bigr]\bigl(-v^{-1}\bigr) - \Upsilon \sum_{n\ge 1}
q^{2n-1}_\pm
\frac{\bar\rho_{n} u_n}{2n-1-\Upsilon}.
\end{equation}
With it, the trans-series for the free energy can be written as
\begin{equation}
\mathcal{F}(h) \sim
-\frac{h^2}{2\pi} \bigl\{u_\ri\times f_v + \CC_{\text{IR}}^\pm \re^{-\frac{2}{\tilde{\alpha}}}{\tilde\alpha}^\Delta\bigr\}.
\label{eq_freeF}
\end{equation}
As before, we define,
\begin{align}
&u_\ri= \sum_{n\geq 0, \ell\geq 0} u_\ri^{(n,\ell)} q^\ell_\pm v^n,\nonumber\\
&f_v= \sum_{n\geq 0, \ell\geq 0} f_v^{(n,\ell)} q^\ell_\pm v^n,\nonumber\\
&\{u_\ri\times f_v\}= \sum_{n\geq 0, \ell\geq 0} \CF_v^{(n,\ell)} q^\ell_\pm v^n.
\end{align}

The above procedure leads to trans-series in $v$. However, we are interested in~trans-series in~the more natural coupling $\tilde{\alpha}$, defined in~\eqref{tilde-alpha-def}.
Using the mass gap for the Gross--Neveu model~\eqref{eq_mass_gap_GN}, we can relate $v$ and $\tilde{\alpha}$ through the equation
\begin{equation}
\biggl(\frac{1}{2v}-\Delta\log v\biggr) - \log u_\ri =\biggl(\frac{1}{\tilde{\alpha}}-\Delta\log\tilde{\alpha}\biggr)+\frac{\Upsilon}{2}\log\Upsilon +\log 2.
\label{v-to-a}
\end{equation}
It is also convenient to write
\begin{equation}
y_\pm = \Upsilon 2^{2/\Upsilon } \left(u_\ri\right)^{2/\Upsilon}q_\pm,
\label{k-to-q}
\end{equation}
where
\begin{equation}
y_\pm = \bigl(\re^{\mp\frac{\ri\pi\Delta}{\Upsilon}}\bigr)\re^{- \frac{2}{\Upsilon \tilde{\alpha}}}(\tilde\alpha)^{\frac{2\Delta}{\Upsilon}}.
\end{equation}
The most convenient way of changing from $v$ to $\tilde{\alpha}$ is to treat $v$ and $q_\pm$ as independent parameters and change variables into $\tilde{\alpha}$ and $y_\pm$ using both~\eqref{v-to-a} and~\eqref{k-to-q}. Let us also introduce the notation
\begin{align}
q_\pm= \sum_{\ell\geq 1, n\geq 0} q_{n,\ell} y^\ell_\pm \tilde\alpha^n,\qquad
\{u_\ri\times f_v\}= \sum_{\ell\geq 0, n\geq 0} a_{n, \ell} y^\ell_\pm \tilde\alpha^n.
\end{align}
With these ingredients we can derive the real coefficients $a_{n, \ell}$, from which we construct $\CG_n(y)$ as in (\ref{CGdef}).
Since the overall phase of the trans-series sectors factors out, it is irrelevant for considerations of convergence.

\subsubsection[Leading orders in tilde{alpha}]{Leading orders in $\boldsymbol{\tilde{\alpha}}$}

The leading order in $\tilde{\alpha}$ is obtained by neglecting all terms with the integral operator $\mathfrak{I}$. In this case, we obtain simple recursive relations. The $u_k$ are solved by
\begin{equation}
u_k^{(0,0)}=\frac{\Upsilon}{2k-1},\qquad
u_k^{(0,s)}=\frac{\Upsilon}{2}\sum_{\tfrac{s+1}{2}\geq r\geq 1} \frac{\bar\rho_r u_r^{(0,s+1-2r)}}{k+r-1},
\end{equation}
which is simple to solve order by order, although hard to solve in closed form.
Similarly, we obtain
\begin{align}
&u_\ri^{(0,0)}=1,\qquad
u_\ri^{(0,s)}=\Upsilon\sum_{\tfrac{s+1}{2}\geq r\geq 1} \frac{\bar\rho_r u_r^{(0,s+1-2r)}}{2r-1+\Upsilon},\nonumber\\
&f_v^{(0,0)}=1,\qquad
f_v^{(0,s)}=-\Upsilon\sum_{\tfrac{s+1}{2}\geq r\geq 1} \frac{\bar\rho_r u_r^{(0,s+1-2r)}}{2r-1-\Upsilon}.
\end{align}
From these building blocks, regular power series manipulations are sufficient to extract $a_{0, \ell}$. For example, we obtain
\begin{align*}
&a_{0, 0}=1,
\qquad
a_{0, 1}=
\frac{2^{\frac{\Upsilon -2}{\Upsilon }} \Upsilon ^2 \bar{\rho }_1}{\Upsilon ^2-1}
,
\qquad
a_{0,2}=
\frac{2^{\frac{\Upsilon -4}{\Upsilon }} \Upsilon ^2 \bar{\rho }_1^2}{(\Upsilon +1)^2}
,\nonumber\\
&a_{0,3}=
\frac{1}{3} 2^{-\frac{\Upsilon +6}{\Upsilon }} \biggl(\frac{3 \Upsilon ^2 (3 \Upsilon -7) \bar{\rho }_1^3}{(\Upsilon +1)^3}+\frac{4 \bar{\rho }_2}{\Upsilon ^2-9}\biggr)
.
\end{align*}
All the $a_{n,\ell}$ are polynomials in~the residues $\bar\rho_i$, but only monomials
$\bar\rho_{i_1}^{p_1}\bar\rho_{i_2}^{p_2}\cdots\bar\rho_{i_k}^{p_k}$ that satisfy
\begin{equation}
\sum_{j=1}^k p_j (2i_j-1) = \ell
\end{equation}
contribute to $a_{n,\ell}$. The polynomial coefficients for $a_{0,\ell}$ are rational up to an
overall irrational power of $2$, and only at higher orders in $\tilde\alpha$ does one find 
transcendental numbers. Nevertheless, the $\bar\rho_k$ are themselves transcendental, as can be seen in~\eqref{polepm}.


The next to leading order in $\tilde{\alpha}$ can be immediately obtain from the leading one. Because the perturbative part of $\bar{u}(\eta)$ is one order lower than the non-perturbative, i.e.,
\be
u^{(0,0)}(\eta)= \frac{1}{\eta} , \qquad u^{(0,n)}(\eta)= 0, \qquad u^{(1,\ell)}(\eta)= \bar{u}^{(\ell)}(\eta)\big|_{v=0},
\ee
the recursive problem is the same and the solution only differs by an overall constant.
Concretely,
\begin{alignat}{3}
&u_k^{(1,s)} = -\Delta u_k^{(0,s)},\qquad&& u_\ri^{(1,s)}=-\Delta u_\ri^{(0,s)},&\nonumber\\
&f_v^{(1,s)}=-\Delta f_v^{(0,s)} \qquad&& \CF_v^{(1,s)}=-2\Delta \CF_v^{(0,s)}.&
\label{lo-to-nlo}
\end{alignat}
For the change of variables, we find first that
\begin{equation}
v = \frac{\tilde{\alpha}}{2}+\CO\bigl(\tilde{\alpha}^2\bigr),
\end{equation}
without any term proportional to $q_\pm$ at order $\tilde{\alpha}$ or lower. Furthermore,
\begin{equation}
q_{1, \ell} = \frac{\Delta \ell}{\Upsilon} q_{0, \ell},
\end{equation}
which leads to
\begin{equation}
a_{1, \ell} = \Delta\biggl(\frac{\ell}{\Upsilon}-1\biggr)a_{0, \ell}.
\label{F1-f0}
\end{equation}
Naturally, $\CG_1(y)$ has the same radius of convergence as $\CG_0(y)$.
At the next order in $\tilde\alpha$, there are no such shortcuts.

\subsubsection{Numerical analysis}

Using these recursion relations, the coefficients $a_{0,\ell}$ can be obtained analytically, order by order. However, their expressions are very intricate and a general closed formula is unreachable.

\begin{figure}[t]
\centering
\includegraphics[width=0.6\textwidth]{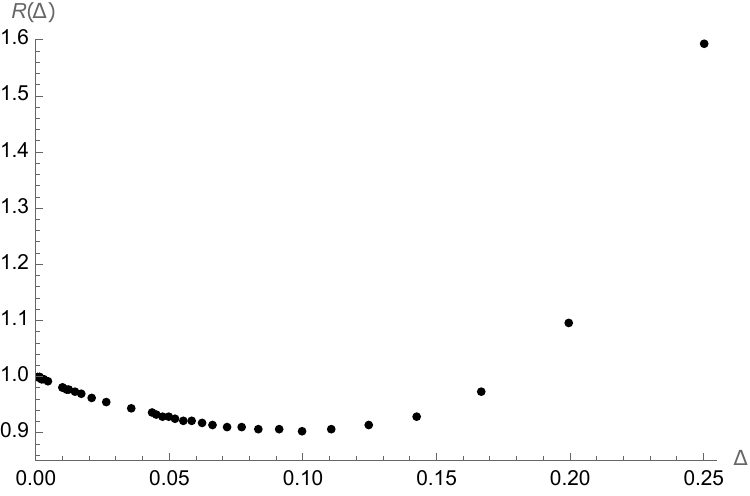}
\caption{Estimates for the radius of convergence $R$ as a function of $\Delta=(N-2)^{-1}$.}\label{rad-gn-plot}
\end{figure}

One straightforward approach is to obtain~the coefficients as exact functions of $N$. This way, one can quickly generate data for different values of $N$, but it takes significant time to compute the series. We calculated 20 exact coefficients in~this form. In a similar strategy to previous cases, we estimated the radius of convergence $R$ with the 17-th coefficient of the second Richardson transform of $a_{0,\ell}/a_{0,\ell+1}$. The convergence error is below 1\%.
In Figure~\ref{rad-gn-plot}, we plot this estimate with all values of $N$ between 6 and 25, as well as from 30 to 100 with steps of 10 and then from 100 to 1000 with steps of 100. In the large $N$ limit we seem to have $R(0)=1$, which agrees with the explicit large $N$ solution, as we explain below. Extrapolating to non-integer values of $N$, the radius seems to diverge at $\Delta=1/3$ ($N=5$). This is consistent with the fact that the trans-series vanishes at $N=5$ (there is only perturbation theory and the IR renormalon).

\begin{table}[t]
\centering
\arraycolsep=3mm
\caption{Estimated radius as a function of $N$, using 200 numeric coefficients. $R$ is the radius in~the $y$-plane, while $R_h$ is the radius in~the $h$-plane. }\label{tabrad}\vspace{2mm}

\begin{tabular}{llll}
\toprule
 $N$ & $R$ & $R_h/m$ & $b_0$\\
 \midrule
 6 & $1.588595196530307661805628(3\pm 7)$ & 0.71482$\dots$ & $2.500000(0 \pm 4)$ \\
 7 & $1.0901795077562836000(83\pm 13)$ & 0.80858$\dots$ & $2.500000(2\pm 7 )$ \\
 8 & $0.969970536544952097(5\pm 4)$ & 0.85994$\dots$ & $2.5000(0\pm 5)$ \\
 9 & $0.92721010523279402(59\pm 35)$ & 0.89194$\dots$ & $2.499(99\pm 10)$ \\
 10 & $0.9101379331724643(0\pm 8)$ & 0.91349$\dots$ & $2.500(0\pm 8)$ \\
 11 & $0.903644442840670(26 \pm 11)$ & 0.92881$\dots$ & $2.50(00\pm 14)$ \\
 12 & $0.90204273932404(39\pm 22 )$ & 0.94016$\dots$ & $2.5(00\pm 22)$ \\
 1000 & $0.997821(23 \pm 14 )$ & 0.99997$\dots$ & $2.4(9\pm 2)$ \\
\bottomrule
\end{tabular}
\end{table}

More efficiently, one can calculate the coefficients numerically from the onset. Taking 200 digits of precision in~the calculation of the $\bar\rho_n$, one can generate 200 coefficients at fixed $\Delta$ in~a~few minutes. We can obtain more precise estimates for the radius this way, see Table~\ref{tabrad}. We estimate the error from the~convergence of successive Richardson transforms. Strikingly, the~convergence is much faster than in~the case of bosonic models. We do not compute the radius of convergence for the subleading series
in~this model, since we already know that it is identical to $R$ due to~\eqref{F1-f0}.

It is also instructive to write the radius in~terms of the variable $h$, which can be related to $y$ through~\eqref{tilde-alpha-def}. The radius in~the $h$-plane, $R_h$, is related to the radius in~the $y$ plane by
\begin{equation}
R_h = m \frac{\Gamma (1-\Delta )}{(2 \re)^{\Delta }} R^{-\frac{1}{2} (1-2 \Delta )}.
\end{equation}
The exact solution $\CF(h)$ is expected to be singular around its zero at $h=m$, hence one could conjecture that the radius of convergence of the trans-series in~the $h$-plane should be $m$. This is not the case for our radius estimate $R_h$, except at large $N$, as can be seen in~Table~\ref{tabrad}. However, as we previously discussed, this radius is simply an approximation of the radius of convergence of the full trans-series.

Lastly, assuming the form~\eqref{an-bn-R},
 we can estimate $b_0$ by evaluating the auxiliary series on the left-hand side of (\ref{bn-seq}) for $n=0$,
where we must plug in~the estimate for the radius in~Table~\ref{tabrad}.
Our results suggest that $b_0=5/2$ and $b_1=3/2$ independently of $\Delta$. Like in~the case of bosonic models, one must account for the error both from the convergence of the sequence and from the estimate of $R$.

\subsubsection[Large N solution]{Large $\boldsymbol{N}$ solution}

It is interesting to compare our results with the large $N$ expansion of the free energy in~\cite{fnw1,fnw2} (see also~\cite{dmss} for further details). There, it is obtained that the leading order term in~the small~$\Delta$ (large $N$) expansion of the free energy,
\begin{equation}
\CF(h) = -\frac{h^2}{2\pi}\sum_{n\geq 0} \CF_n(h) \Delta^{n},
\end{equation}
is given by
\begin{equation}
\CF_0(h) = \sqrt{1-\frac{m^2}{h^2}}-\frac{m^2}{h^2} \cosh ^{-1}\biggl(\frac{h}{m}\biggr).
\end{equation}
As expected, $\CF_0(m)=0$. Furthermore, in~the $h$-plane, there is a branch cut at $[-m,m]$ due to the square root,
and another along $(-\infty,m]$ due to the logarithmic behavior of $\cosh^{-1}$. We see in particular the singularity at
$h=m$ shown in Figure~\ref{fig-singfh}.

In order to make contact with the finite $N$ analysis, it is convenient to re-express $\CF_0(h)$ in~terms of $\tilde\alpha$ and obtain its trans-series expansion. To the first few orders we obtain
\begin{equation}
\CF_0(h)\sim
1
-\re^{-2/\tilde\alpha } \biggl(\frac{1}{\tilde\alpha }+\frac{1}{2}+\log (2)\biggr)
+\frac{\re^{-4/\tilde\alpha }}{8}
+\frac{\re^{-6/\tilde\alpha }}{32}
+\cdots,
\end{equation}
where further terms in $\re^{-2n/\tilde\alpha}$ have rational coefficients and no perturbative corrections.
Putting aside the leading terms, which account for the logarithmic branch cut at $\infty$ in~the $h$-plane, we can treat this trans-series as a convergent series in $\re^{-2/\tilde\alpha}$ with radius of convergence $1$, as can be quickly inspected from the position of the remaining branch cut \big(roughly, $\re^{-2/\tilde{\alpha}}\propto (m/h)^2$\big). This matches the radius $R(\Delta\rightarrow 0)=1$ obtained numerically above.
The truncation of perturbative series does not happen for $\CF_1(h)$, which has also been computed analytically, neither does it happen in subsequent terms.

As a check of consistency, one can also take the finite $N$ trans-series terms obtained with the recursive method and expand them for small $\Delta$. As discussed in~\cite{mmr-an}, this moves the non-perturbative effects $\exp\bigl(-\frac{2}{\Upsilon\tilde\alpha}\bigr)\rightarrow\exp\bigl(-\frac{2}{\tilde\alpha}\bigr)$ and leads to the violation of ``strong resurgence'' observed in~\cite{dmss}. Nevertheless, term by term, the two expansion match, which has been tested beyond 15-th order in $\re^{-2/\tilde\alpha}$ (and NLO in $\tilde{\alpha}$ and $\Delta$).

\section{Conclusions and open problems}

In this paper we have addressed a question which we think is important in~the
resurgence program of QFT: when is the infinite sum over
the non-perturbative sectors well defined, after Borel resummation in~the coupling constant?
Inspired by~\cite{costin-costin}, we have addressed a proxy question, concerning the
convergence of the infinite family of what we have called partial series. In some simple cases
one can establish this property relatively easily, either numerically (as~in~the example of large $N$
sigma models) or analytically (in complex CS theory). We conjecture that this convergence
property is also true for the trans-series of the free energy in integrable asymptotically free theories. Our main effort
in~this paper has been to provide partial but non-trivial
positive evidence for this conjecture, in Section~\ref{sec-integrable}.

When this convergence property fails, as it has been argued to be the case for generic
correlation functions in many QFTs, one has to extend the simple resurgent framework used
in~the examples in~this paper, and
introduce, e.g., a further resummation of the infinite series of non-perturbative sectors, and maybe
more general trans-series. This is the well-known problem of the ``divergence of the OPE" pointed out in, e.g.,
\cite{shifman,shifman-hadrons}, which has been studied in~the phenomenological literature
(see, e.g.,~\cite{peris} for a summary and references). We believe that the convergence property, when it holds, is at the heart of the
successful examples of resurgence in QFT obtained in complex CS theory and in integrable models.

One corollary of our analysis is that the free energy of integrable field theories is a
very special observable from the point of resurgence,
and this might be related to a simpler singularity structure, as compared to generic
correlation functions. As we mentioned in Section~\ref{sec-integrable}, $\CF(h)$ has a
singularity at $h=m$ (see Figure~\ref{fig-singfh}), and it might well
happen that this is the only singularity in~the complex plane of $h$. Given our results on the
convergence of partial series,
we shouldn't have an unbounded
sequence of singularities, as in a finite $N$ two-point function. Clarifying and understanding the global analytic structure of the
free energy is then an important problem for the future. Our results also suggest that the generalization
of this free energy to non-integrable theories might lead to a QFT observable with better resurgent properties.

Clearly, there are many aspects of our work that can be improved. A direct proof, or at least more thorough verification,
of the convergence of the trans-series (\ref{br-transseries}) in integrable field theories would be desirable and reassuring.
The recent progress in~\cite{bbv2} might make this a realistic goal, and it can be probably used to test the convergence
of partial series for higher values of $n$.

The free energy of integrable asymptotically free theories has been an excellent
laboratory to better understand trans-series in QFT. It has confirmed many expectations
but also led to surprises, like the appearance of new renormalon
singularities in~the Borel plane~\cite{mmr-an}. Similarly, it would be very
interesting to find a workable, exactly solvable model displaying the divergence of the OPE. There have been
proposals for toy models with divergent trans-series~\cite{shifman, shifman-hadrons, peris},
but they do not really emerge from an explicit exact solution to a QFT. Such an exact solution
would provide a sounder foundation for phenomenological models and might clarify the resurgent structure in~this more complicated setting.

Since the pre-publication of the manuscript of this paper, there have been developments in~the literature concerning the convergence of trans-series in QFT. An example of a tractable divergent OPE trans-series with many of the characteristic described above has been analyzed in~\cite{Liu:2025edu}. Meanwhile, a new example of a convergent trans-series in~the context of large $N_f$ QED has been studied in~\cite{Laenen:2023hzu}. In the realm of the integrable models studied in~this paper, there has also been significant developments in~\cite{Bajnok:2025mxi}. Using the techniques from~\cite{bbv2}, the authors obtain~the non-perturbative sectors of ${\rm O}(N)$ NLSM to high perturbative order. They verify that the partial series are convergent with identical radius of convergence, using terms up to perturbative order~10. Furthermore, they perform the Borel summation of each sector and verify that indeed this radius corresponds to the radius of convergence of the sum over the Borel summed series. Thus, they numerically validate the analysis started in~this paper.

\subsection*{Acknowledgements}
We would like to thank Zoltan Bajnok, Janos Balog, Ovidiu Costin,
Stavros Garoufalidis, Santiago Peris, Marco Serone and Istvan Vona for useful comments
and discussions. The work of M.M.\ and R.M.\ has been supported in part by the ERC-SyG project
``Recursive and Exact New Quantum Theory'' (ReNewQuantum), which
received funding from the European Research Council (ERC) under the European
Union's Horizon 2020 research and innovation program,
grant agreement No. 810573.
The work of T.R.\ is supported by the ERC-COG grant NP-QFT No.~864583
``Non-perturbative dynamics of quantum fields: from new deconfined
phases of matter to quantum black holes'', by the MUR-FARE2020 grant
No. R20E8NR3HX ``The Emergence of Quantum Gravity from Strong Coupling
Dynamics'',
and by INFN Iniziativa Specifica GAST and ST\&FI.
We would also like to thank the SwissMAP Research Station at Les Diablerets for hosting the conference
``Resurgence and quantization'', which allowed us to discuss the results of this paper with many colleagues.

\pdfbookmark[1]{References}{ref}
\LastPageEnding

\end{document}